\documentclass[12pt,a4paper]{article}

\usepackage[
a4paper,
left=0.9in,
right=0.9in,
top=0.75in,
bottom=0.75in
]{geometry}

\usepackage{amsmath,amssymb,amsfonts}
\usepackage{bm}
\usepackage{newtxtext,newtxmath}

\numberwithin{equation}{section}

\usepackage{graphicx}
\usepackage{subcaption}
\usepackage{float}
\usepackage[percent]{overpic}
\usepackage[numbers,sort&compress]{natbib}

\usepackage[
colorlinks=true,
linkcolor=blue,
citecolor=blue,
urlcolor=blue
]{hyperref}
\begin{document}
\begin{center}
{\large\bfseries Electrophoretic motion of a liquid droplet with Brinkman-screened internal hydrodynamics\par}
		
		\vspace{0.8em}
		
		{\small Sutapa Mandal$^{1,\dagger}$ and Subrata Majhi$^{1,\dagger,*}$\par}
		
		\vspace{0.5em}
		
		{\footnotesize
			$^{1}$Department of Mechanics and Aerospace Engineering,\\
			Southern University of Science and Technology, Shenzhen 518055, Guangdong, P.R. China\\[0.35em]
			\par}
	\end{center}
		
	\begin{center}
	\begin{minipage}{1\textwidth}
	\small
We develop a theory for the electrophoresis of a spherical porous liquid droplet with prescribed uniform surface charge. The exterior electrokinetics is governed by the Poisson--Nernst--Planck--Stokes equations, while the internal liquid motion is described by the Brinkman--Debye--Bueche equation. A regular perturbation expansion in the applied electric field reduces the governing equations to coupled radial ordinary differential equations. In the Debye--H\"uckel regime, we derive a closed-form mobility expression valid for arbitrary Debye layer thickness. The analysis shows that the porous interior modifies clean-droplet electrophoresis through a single Brinkman-screened hydrodynamic resistance, yielding a continuous transition between clean-droplet and rigid-particle limits. Numerical solutions beyond the low-potential regime reveal a non-universal role of permeability: increasing the Darcy number can either suppress or enhance the mobility. This reversal is determined by the sign of the interfacial-velocity mode, which is governed by the competition between tangential Maxwell traction and hydrodynamic shear generated by electric-double-layer distortion. Dielectric polarization, surface charge and double-layer thickness can reverse the internal circulation, while the Darcy number controls how strongly this circulation is transmitted through the porous interior. This permeability sensitivity is especially pronounced for highly polarizable droplets in the thin-double-layer regime. The theory provides a basis for tuning electrokinetic transport of soft porous droplets in microfluidic and biomedical technologies.
\end{minipage}
	\end{center}
		
\begin{center}
		\begin{minipage}{1\textwidth}
		\end{minipage}
	\end{center}
	
	\begingroup
	\renewcommand{\thefootnote}{\fnsymbol{footnote}}
	\footnotetext[2]{These authors contributed equally to this work.}
	\footnotetext[1]{Corresponding author: \texttt{majhisubrata1996@gmail.com}; \texttt{subrata@sustech.edu.cn}.}
	\endgroup
\section{\textbf{Introduction}}\label{Introduction}
Electrophoretic transport of liquid droplets is central to the manipulation and characterization of emulsions, bubbles and soft colloidal inclusions in microfluidic, environmental and biomedical systems \cite{rashidi2021mechanistic}. Unlike rigid colloids, a liquid droplet possesses a mobile fluid--fluid interface, so that its electrophoretic response is governed not only by the electric double layer in the surrounding electrolyte but also by the coupling between interfacial stresses and internal circulation. Electric stresses acting near the charged interface, together with viscous stresses in the two fluids, can generate interfacial motion and vortical flow inside the droplet. Consequently, droplet electrophoresis differs fundamentally from classical particle electrophoresis: mobility is affected by the internal hydrodynamic response of the droplet and cannot, in general, be interpreted solely in terms of the equilibrium surface charge or $\zeta$-potential. A predictive theory must therefore account for both the electrokinetic forcing in the exterior electrolyte and the mechanical response of the droplet interior.

Theoretical studies of droplet electrophoresis have traditionally considered a clean liquid interior governed by the Stokes equations. Booth first derived Debye--H\"uckel mobility expressions for charged liquid droplets by accounting for the continuity of velocity and stress across a spherical fluid--fluid interface \cite{booth1951cataphoresis}. This framework was subsequently extended to conducting mercury drops by Ohshima, Healy \& White \cite{ohshima1984electrokinetic}, and to more general conducting and non-conducting fluid globules with interfacial transport effects by Baygents \& Saville \cite{baygents1991electrophoresis}. More recent studies have clarified the roles of dielectric polarization, tangential Maxwell traction and nonlinear electric-double-layer distortion in the electrophoresis of fluid droplets and bubbles \cite{schnitzer2014electrophoresis,wu2021electrophoresis,mahapatra2022electrophoresis,tsai2022electrophoresis}. These works demonstrate that the mobility of a charged droplet is controlled not only by the equilibrium surface charge or $\zeta$-potential, but also by the manner in which electric stresses and hydrodynamic shear determine interfacial motion and internal circulation.

The interfacial charge condition itself is a subtle modelling issue. In many emulsions and bubbles the charge is associated with adsorbed ions or surface-active species, so that the interface may support lateral charge transport, adsorption--desorption kinetics and surface-tension gradients. These mechanisms can strongly modify the electrophoretic response, because redistribution of interfacial charge changes the tangential Maxwell traction, while non-uniform surfactant concentration generates Marangoni stresses \cite{hill2020dynamic,hill2020electrokinetic,hill2025roles}. Recent fluid-sphere theories have shown that interfacial charge mobility and kinetic exchange may qualitatively alter the mobility and regularize behaviours predicted by idealized constant-charge models \cite{hill2025roles,majhi2025electrophoresis}.

Porous and gel-like droplets constitute an important class of soft dispersed objects in microfluidics and biomedical engineering. In droplet-based platforms, aqueous polymer, monomer, alginate, agarose or hydrogel precursor solutions can be emulsified and subsequently crosslinked to produce hydrogel microdroplets or microgels with tunable size, permeability and mechanical response \cite{zhu2017hydrogel,wan2012microfluidic,orabi2023emerging,daly2020hydrogel,mohamed2020microfluidics,headen2014microfluidic}. Such objects are used as cell-laden microreactors, single-cell analysis compartments, three-dimensional culture environments, microcapsules for controlled release, and carriers for therapeutic or biological cargo. Their importance arises from the fact that the internal polymer network is liquid-saturated and porous, allowing molecular transport through the matrix while providing a mechanically soft environment for encapsulated cells, solutes or active cargo. Their electrohydrodynamic response, however, cannot in general be inferred from that of a clean liquid droplet. A clean droplet supports an internal Stokes circulation driven by interfacial stresses, whereas a porous or gel-like droplet contains a cross-linked polymer network that permits flow of the saturating liquid but resists relative motion between the liquid and the skeleton. Thus, the interfacial motion generated during electrophoresis is transmitted into the droplet interior through a porous-medium hydrodynamic response whose strength is controlled by the permeability.

This observation identifies a gap within the existing literature on droplet electrophoresis. As noted above, classical and modern theories of droplet electrophoresis usually treat the interior as a clean Newtonian liquid  \cite{booth1951cataphoresis,ohshima1984electrokinetic,baygents1991electrophoresis,wu2021electrophoresis,mahapatra2022electrophoresis,tsai2022electrophoresis}. By contrast, electrokinetic theories of soft or porous colloids usually describe polymer-coated particles, gel particles or porous spheres, in which the surrounding electrolyte penetrates a charged polymeric region and the Brinkman--Debye--Bueche resistance appears as a distributed hydrodynamic drag within the permeable phase \cite{hill2005exact,uematsu2015electrophoresis,gopmandal2017effect,ohshima2019gel}. Such systems do not involve an immiscible liquid interior separated from the electrolyte by a mobile fluid--fluid interface, and therefore do not require the droplet-type tangential stress balance coupling internal hydrodynamic stress to exterior hydrodynamic and electric stresses. Recent work on internally resistive droplets has introduced Brinkman hydrodynamics into droplet-migration problems, but the driving considered there is hydrodynamic or thermocapillary rather than electrokinetic \cite{sharma2026thermocapillary}. The electrophoresis of a porous liquid droplet therefore remains a distinct unresolved problem: the interface is mobile and charged as in a liquid droplet, while the internal circulation is Brinkman-screened by the Darcy resistance of the porous matrix. The resulting motion is governed by the coupling between electric-double-layer forcing, interfacial electric traction, hydrodynamic shear and internal porous resistance.

Motivated by this gap, we study the electrophoresis of a spherical porous liquid droplet whose interface carries a prescribed uniform surface charge. The porous skeleton is assumed to be uncharged and impermeable to ions, so that the electric double layer is formed only in the surrounding electrolyte. The exterior electrokinetics is described by the Poisson--Nernst--Planck--Stokes equations, while the electric potential inside the droplet satisfies the Laplace equation and the internal liquid motion is governed by the Brinkman--Debye--Bueche equation. The interface remains spherical, with continuous tangential velocity and a tangential stress balance coupling the internal and external hydrodynamic shear stresses to the tangential Maxwell traction associated with the fixed surface charge. Within a linear-response expansion in the imposed electric field, the governing equations are reduced to a coupled system of radial perturbation equations, which is solved analytically in the Debye--H\"uckel limit and numerically through the finite element method beyond the low-potential regime. 

The analysis reveals that the porous interior does not modify electrophoresis by changing the electric double layer directly. Instead, permeability modifies the hydrodynamic transmission of interfacial motion into internal circulation. In the Debye--H\"uckel limit, this effect appears through a permeability-dependent Brinkman resistance that replaces the internal Stokes resistance of a clean droplet. This gives a continuous hydrodynamic connection between the classical liquid-droplet and rigid-particle limits. Beyond the low-potential regime, the numerical solutions show that the effect of permeability has no universal direction: increasing the Darcy number may either reduce or enhance the mobility, depending on the interfacial-velocity mode determined by the relative contributions of hydrodynamic shear and electric traction. Dielectric polarization, surface charge and double-layer thickness can all alter this mode.

The paper is organized as follows. Section~\ref{Model_description} formulates the electrokinetic model for a porous droplet. Section~\ref{sec:perturbation} develops the weak-field perturbation framework and derives the coupled radial boundary-value problem. Section~\ref{sec:DH_solution} obtains the Debye--H\"uckel mobility and discusses the limiting cases. Section~\ref{numerical_method} describes the numerical implementation in COMSOL Multiphysics. Section~\ref{Discussions} presents results beyond the low-potential regime, with emphasis on the roles of permeability, dielectric polarization, surface charge and double-layer thickness. Section~\ref{Conclusion} summarizes the main conclusions and outlines possible extensions.

\section{\bf Model description}\label{Model_description}
We consider the electrophoretic motion of a spherical porous liquid droplet of radius $a$ and dielectric permittivity $\varepsilon_p$, suspended in an incompressible Newtonian electrolyte solution of dielectric permittivity $\varepsilon_e$, viscosity $\eta$, and density $\rho$. The electrolyte contains monovalent ionic species associated with salts such as NaCl or KCl. The suspension is subjected to a uniform DC electric field $\bm{E}_{\infty}=E_{\infty}\bm{e}_z$, imposed in the far-field electrolyte. The problem is axisymmetric about the $z$-axis. The droplet therefore translates along the field axis with electrophoretic velocity $\bm{U}^{*}_E=U^{*}_E\bm{e}_z$, where the sign of $U^{*}_E$ determines whether the motion is parallel or antiparallel to the imposed field. Quantities carrying a superscript asterisk are dimensional. A spherical polar coordinate system $(r,\theta,\varphi)$ is adopted with its origin at the droplet centre, and the polar axis $(\theta=0)$ is chosen along the direction of the imposed electric field.

The droplet interior is modelled as an uncharged, ion-impermeable porous liquid phase. The liquid contained within the porous droplet flows through a uniformly distributed polymer network, which is represented using the Brinkman--Debye--Bueche continuum description \cite{brinkman1949calculation,debye1948intrinsic}. In this framework, the polymer segments act as resistance centres and exert a frictional drag on the liquid flowing inside the droplet. 
\begin{figure}[h]
	\centering
	\includegraphics[width=4.5in]{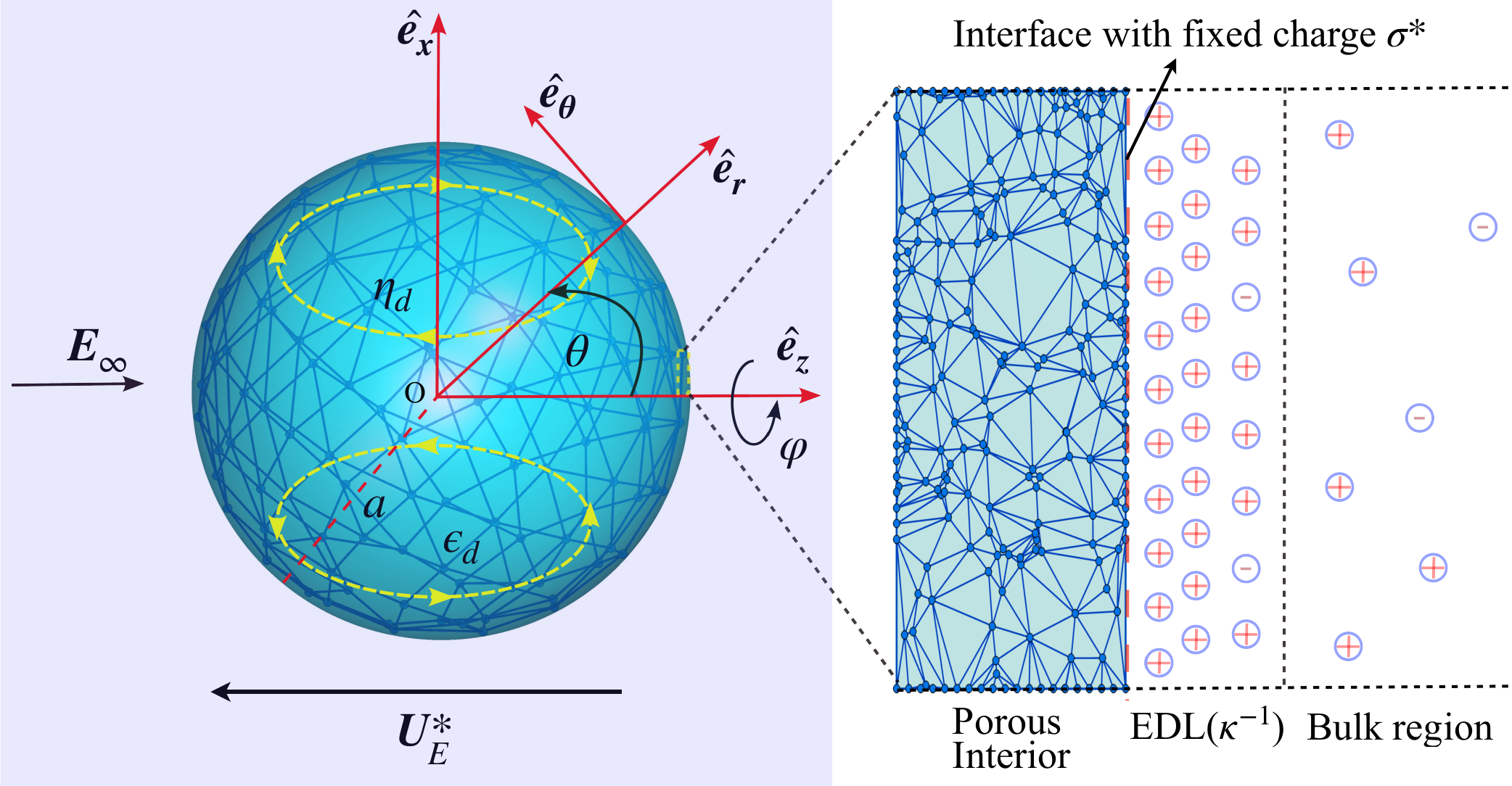}
	\caption{Schematic representation of the electrophoresis of a porous liquid droplet with prescribed negative surface charge density $\sigma^{*}<0$. The droplet translates with electrophoretic velocity $\bm{U}^{*}_E$ under a uniform DC electric field $\bm{E}_\infty=E_\infty\bm{e}_z$. The zoomed view illustrates the ion-free porous interior, the fixed charged interface, the electric double layer (EDL) and the electroneutral bulk electrolyte.}
	\label{fig1}
\end{figure}
The hydrodynamic response of the porous droplet is therefore governed by a Brinkman momentum balance, in which viscous diffusion of momentum in the saturating liquid competes with Darcy resistance exerted by the polymer network:
\begin{subequations}\label{eq:1}
	\begin{gather}
		\label{eq:1a}
		\eta_{d}\nabla^{*}\times\nabla^{*}\times \bm{u}_{d}^{*}(\bm r^*)+\nabla^{*}p_{d}^{*}(\bm r^*)+\frac{\eta_{d}}{K}\bm{u}_{d}^{*}(\bm r^*)=0,\\
		\label{eq:1b}
		\nabla^{*}\cdot\bm{u}_{d}^{*}(\bm r^*)=0,
	\end{gather}
\end{subequations}
Here $\bm{u}_{d}^{*}(\bm r^*)$ and $p_d^{*}(\bm r^*)$ denote the dimensional velocity and pressure fields at a position $\bm r^*$ inside the porous droplet. The last term in Eq.~\eqref{eq:1a} represents the hydrodynamic resistance exerted by the porous polymer network on the liquid, where $K$ is the Darcy permeability of the internal porous structure. The associated Brinkman screening length is $\ell=\sqrt{K}$. Since the porous skeleton is assumed to be uncharged and mobile ions are absent inside the droplet, no electrical body-force term appears in Eq.~\eqref{eq:1a}. In the present study, the viscosity ($\eta_d$) of the liquid saturating the porous droplet is taken to be constant. Although an effective viscosity may be introduced to account for additional viscous stresses associated with detailed pore-scale structure, such a description requires a constitutive relation depending on the pore geometry and porosity. We therefore retain the minimal Brinkman--Debye--Bueche model in order to isolate the influence of permeability on the electrophoretic response of the porous droplet. Using $a$, $U_0=\varepsilon_e\phi_0^2/(\eta a)$ and $\varepsilon_e\phi_0^2/a^2$ as the length, velocity and pressure scales, respectively, Eqs.~\eqref{eq:1a}--\eqref{eq:1b} become
\begin{subequations}\label{eq:2}
	\begin{gather}
		\label{eq:2a}
		\eta_r\nabla\times\nabla\times\bm{u}_d(\bm{r})+\nabla p_d(\bm{r})+\frac{\eta_r}{Da}\bm{u}_d(\bm{r})=0,\\
		\label{eq:2b}
		\nabla\cdot\bm{u}_d(\bm{r})=0,
	\end{gather}
\end{subequations}
The dimensionless velocity field inside the droplet is written as $\bm{u}_d=v_d\,\bm{e}_r+u_d\,\bm{e}_\theta$, where $v_d$ and $u_d$ are the radial and polar velocity components, respectively. Here $\eta_r=\eta_d/\eta$ is the viscosity ratio between the liquid inside the droplet and the external electrolyte. The parameter $Da=K/a^2$ is the Darcy number, where $K$ is the Darcy permeability of the porous skeleton. Thus, $Da$ compares the intrinsic permeability length scale $\sqrt{K}$ with the droplet radius $a$. Physically, it measures how easily the liquid inside the porous droplet can flow through the polymer network. On the other hand, the motion of the incompressible Newtonian electrolyte outside the droplet can be expressed in dimensionless form as
\begin{subequations}\label{eq:3}
	\begin{gather}
		\label{eq:3a}
		\nabla p(\bm{r})+\nabla\times\nabla\times \bm{u}(\bm{r})+ \frac{(\kappa a)^{2}}{2} \rho_e(\bm{r}) \nabla \phi(\bm{r})=0,\\
		\label{eq:3b}
		\nabla \cdot \bm{u}(\bm{r})=0,
	\end{gather}
\end{subequations}
where $\bm{u}=(v,u)$ is the velocity vector, $v$ is the radial and $u$ is the polar velocity components, $p$ is the pressure and $\phi$ is the electric potential at a position $\bm{r}$ outside the droplet. The electric potential is scaled by $\phi_0$. The dimensionless space charge density is $\rho_e(\bm{r})=\sum_{i=1}^{N}{z_{i}n_{i}(\bm{r})}$, scaled by $eI$. $n_{i}(\bm{r})$ is the concentration of the i$^{th}$ ionic species at $\bm{r}$ and $z_{i}$ is its valency. The quantity $\kappa^{-1}$ is the Debye length defined as $\kappa^{-1}=\sqrt{\varepsilon_e\phi_{0}/2Ie}$ and $I$ is the ionic strength at the bulk defined by $I=(1/2)\sum_{i=1}^{N}{z^{2}_{i}n^{\infty}_{i}}$, where $n^{\infty}_{i}$ is the bulk concentration of i$^{th}$ ionic species. The inertial terms in the momentum equations (\ref{eq:2a},\ref{eq:3a}) are neglected because the Reynolds numbers associated with the internal and external flows are small, $Re, Re_{d}\sim O(10^{-4})$. The boundary conditions for the velocity field at the droplet surface ($r=1$) are the tangential velocity continuity, fluid impermeability and the stress balance conditions \cite{leal2007advanced} i.e.,
\begin{subequations}\label{eq:4}
	\begin{gather}
		\label{eq:4a}
		\bm{u}=\bm{u}_{d},\\
		\label{eq:4b}
		\bm{n}\cdot\bm{u}=0,\\
		\label{eq:4c}
		\bm{n}\cdot(\bm{S}_{H}+\bm{S}_{E})-Ca^{-1}\gamma(\nabla_s\cdot\bm{n})\bm{n}+Ca^{-1}\nabla_{s}\gamma=\bm{n}\cdot(\bm{S}_{d,H}+{\bm{S}}_{d,E})
	\end{gather}
\end{subequations}
Here, $\bm S_{H}$ and $\bm S_{d,H}$ are the dimensionless hydrodynamic stresses outside and inside the droplet respectively, i.e., $\bm S_{H}=-p\boldsymbol{I}+[\nabla \bm{u} + (\nabla \bm{u})^T]$ and $\bm S_{d,H}=-p_{d}\boldsymbol{I}+\eta_{r}[\nabla \bm{u}_{d} + (\nabla \bm{u}_{d})^T]$. $\bm S_{E}$ and $\bm S_{d,E}$ are the scaled Maxwell stresses and are defined as $\bm S_{E}=\left[\boldsymbol{EE}-(1/2){E}^2 \boldsymbol{I}\right]$ and $\bm S_{d,E}=\varepsilon_{r}\left[\boldsymbol{E_dE_d}-(1/2){E_d}^2 \boldsymbol{I}\right]$. Here $\boldsymbol{E}=-\nabla \phi$, $(E)^2=\boldsymbol{E\cdot E}$ and $\boldsymbol{EE}$ denotes the dyadic product. The parameter $\varepsilon_{r}$ represents the dielectric permittivity ratio between the droplet and the electrolyte medium. The operator $\nabla_{s}=(\boldsymbol{I}-\bm{n}\bm{n})\cdot\nabla$ is the gradient along the surface. The variable $\gamma=\gamma^*/\gamma_0$ denotes the dimensionless surface tension, where $\gamma_0$ is the interfacial tension in the absence of surfactants, and $Ca=\varepsilon_e\phi_0^2/(\gamma_0 a)$ is the capillary number based on $\gamma_0$. As discussed in Section \ref{Introduction}, for charged surfactant-laden droplets, interfacial charge mobility, adsorption--desorption kinetics and Marangoni stresses that arise due to the non-uniform interfacial surfactant concentration, can substantially modify electrophoretic motion. In the present study, we deliberately exclude these effects by assuming a spatially uniform surfactant distribution at the interface. Consequently, the surface tension is uniform, so that $\nabla_s\gamma=0$, and the surface charge density is prescribed and fixed. This assumption is consistent with the experimental strategy of Yang, Shin \& Stone \cite{yang2018diffusiophoresis}, who minimized the surfactant-concentration gradient by maintaining the same bulk SDS concentration in the droplet suspension and in the initial high-salt solution. Thus, tangential stress balance condition can be obtained upon doing the algebraic simplification in Eq.(\ref{eq:4c}) as
\begin{equation}\label{eq:5}
	\left(\frac{\partial u}{\partial r}-\frac{u}{r}\right)_{r=1^+}-\eta_r \left(\frac{\partial u_d}{\partial r}-\frac{u_d}{r}\right)_{r=1^-} =\sigma\frac{\partial \phi}{\partial \theta}\bigg|_{r=1}.
\end{equation} 
The left-hand side of the Eq.(\ref{eq:5}) is the jump in hydrodynamic shear stress across the interface, whereas the right-hand side is the corresponding tangential Maxwell traction contribution exerted by the fixed surface charge. The normal stress balance is not used to determine the electrophoretic velocity because the droplet is assumed to remain spherical. In the weak-field regime considered here, deformation induced by the imposed electric field enters at $O(\Lambda^2)$ \cite{taylor1964disintegration,torza1971electrohydrodynamic,baygents1991electrophoresis}, whereas the present perturbation analysis is carried out only to $O(\Lambda)$. Here, $\Lambda$ is the dimensionless imposed electric field, defined later in this section. The normal stress condition therefore only determines the pressure jump required to maintain the spherical interface and does not affect the leading-order mobility \cite{schnitzer2014electrophoresis,majhi2025electrophoresis}. The dimensionless surface charge density is defined as $\sigma={\sigma^*a}/{\varepsilon_e\phi_0}$, where $\sigma^*$ is the dimensional surface charge density. As described before, the surface charge density is considered to be fixed.

The governing equation for the electric potential $\phi(\bm{r})$ outside the droplet is obtained from the Gauss's law as 
\begin{equation}\label{eq:6}
	\nabla^{2}\phi(\bm{r})=-\frac{(\kappa a)^2}{2}\sum_{i=1}^{N}{z_{i}n_{i}(\bm{r})}
\end{equation}
and the electric potential inside the droplet i.e., $\phi_{d}(\bm{r})$ is governed by the Laplace equation as there are no free ions inside the droplet and thus it is given as 
\begin{equation}\label{eq:7}
	\nabla^{2}\phi_{d}(\bm{r})=0.
\end{equation}
The boundary conditions for the electric potential at the droplet surface $(r=1)$ are 
\begin{equation}\label{eq:8}
	\bm{n}\cdot\left(\nabla\phi-\varepsilon_{r} \nabla \phi_{d} \right)=-\sigma~~~\text{and}~~~\phi=\phi_{d}.
\end{equation}
The conservation law for the flux due to the $i^{th}$ ionic species is
\begin{equation}
	\nabla\cdot[Pe_{i}n_{i}(\bm r)\bm{u}(\bm r)-n_{i}(\bm r)\nabla \mu_{i}(\bm r)]=0,
	\label{eq5}
\end{equation}
where $Pe_i=U_0a/D_i$ is the ionic Péclet number of the $i$th species and $\mu_{i}(\bm r)$ is the dimensionless electrochemical potential given by 
\begin{equation}
	\mu_{i}(\bm r)=\mu_i^{\rm ref}+z_{i}\phi(\bm r)+\ln{n_{i}(\bm r)}.
	\label{eqAAA}
\end{equation}
Here, $\mu_i^{\rm ref}$ is the reference value of the electrochemical potential. In the present study, ions are treated as point charges and therefore, non-ideal excess electrochemical potential contributions are neglected. The boundary conditions for the ion concentration at the droplet surface ($r=1$) is the ion impermeability condition, which is $\left[Pe_i n_i\bm{u}-n_i\nabla\mu_i\right]\cdot\bm{e}_r=0$ at $r=1$. Since the interface is impermeable to fluid motion, $v(1)=0$, this reduces to
\begin{equation}
	\nabla\mu_i\cdot\bm{e}_r=0
	\qquad \text{at } r=1.
	\label{eq6}
\end{equation}

In the laboratory frame the droplet translates with velocity $U_E\bm{e}_z$, while in the droplet-fixed frame the undisturbed electrolyte approaches the droplet with velocity $-U_E\bm{e}_z$. Thus, the far-field conditions in the droplet-fixed reference frame are
\begin{equation}
	\bm{u}\rightarrow -U_E\bm{e}_z,\qquad
	\phi\sim-\Lambda r\cos\theta,\qquad
	n_i\rightarrow n_i^\infty/I
	\quad \text{as}\quad r\rightarrow\infty,
	\label{eq12}
\end{equation}
where $\Lambda=E_\infty a/\phi_0$ is the dimensionless imposed electric field. The electrophoretic velocity $U_E$ is determined from the force-free condition on any arbitrary surface ${S}$ enclosing the droplet \cite{schnitzer2014electrophoresis,cobos2023nonlinear},
\begin{equation}\label{eq:forcefree}
	\oint_{S}\left(\bm S_H+\bm S_E\right)\cdot\bm{n}\,dS=0.
\end{equation} 
\section{\bf Weak-field perturbation analysis} \label{sec:perturbation}
We now develop a regular perturbation formulation for the electrophoretic motion of the porous droplet in the weak-field regime. In typical electrophoretic experiments involving colloidal particles or droplets \cite{baygents1991electrophoresis,marinova1996charging,tottori2019nonlinear}, the applied electric field is sufficiently weak such that $\Lambda\ll1$. Under this condition, the electrostatic, ionic concentration and hydrodynamic fields may be expanded as regular perturbations about their equilibrium states in the form
\begin{equation}
	\mathcal{Q}(\bm r)=\mathcal{Q}^{0}(r)+\Lambda\mathcal{Q}^{(1)}(r,\theta)+O(\Lambda^{2}),\qquad \mathcal{Q}=\begin{bmatrix}\phi&\phi_d&n_i&\mu_i&\bm{u}&\bm{u}_d\end{bmatrix}^{T},\quad i=1,\ldots,N .
	\label{eq:general_perturbation_expansion}
\end{equation}
Here $\mathcal{Q}^{0}$ denotes the equilibrium state and $\mathcal{Q}^{(1)}$ denotes the field-induced perturbation. More explicitly, \(\mathcal{Q}^{0}=\begin{bmatrix}\phi^0&\phi_d^0&n_i^0&\mu_i^{0,\mathrm{eq}}&\bm{0}&\bm{0}\end{bmatrix}^{T}\) and \(\mathcal{Q}^{(1)}=\begin{bmatrix}\phi^{(1)}&\phi_d^{(1)}&n_i^{(1)}&\mu_i^{(1)}&\bm{u}^{(1)}&\bm{u}_d^{(1)}\end{bmatrix}^{T}\). The expansion \eqref{eq:general_perturbation_expansion} separates the governing equations into an \(O(1)\) equilibrium problem and an \(O(\Lambda)\) electrophoretic response.
\subsection{Equilibrium double-layer structure}
At leading order, the electrical double layer is spherically symmetric and there is no fluid motion either inside or outside the droplet. Thus, $	\bm{u}^{0}=\bm{0}$ and $\bm{u}^{0}_{d}=\bm{0}$. The equilibrium electrochemical potential of each ionic species is spatially uniform. Hence $z_i\phi^0(r)+\ln n_i^0(r)=\mathrm{constant}$. Choosing the constant so that \(n_i^0\to n_i^\infty/I\) as \(r\to\infty\), we obtain the Boltzmann distribution
\begin{equation}
	n_i^0(r)=\frac{n_i^\infty}{I}\exp[-z_i\phi^0(r)] .
	\label{eq:equilibrium_boltzmann}
\end{equation}
Substitution of Eq.~\eqref{eq:equilibrium_boltzmann} into the exterior Poisson equation \eqref{eq:6} gives the equilibrium Poisson--Boltzmann equation
\begin{equation}
	\frac{1}{r^2}\frac{d}{dr}\left(r^2\frac{d\phi^0}{dr}\right)=-\frac{(\kappa a)^2}{2}\sum_{i=1}^{N}z_i n_i^0(r),\qquad r>1 .
	\label{eq:equilibrium_PB}
\end{equation}
The corresponding boundary condition at the interface follows from the normal electric-displacement condition in Eq.~\eqref{eq:8}. Since the equilibrium interior potential satisfies the Laplace equation \eqref{eq:7} and is bounded and spherically symmetric, $\phi_d^0$ is constant and hence $d\phi_d^0/dr=0$. The fixed surface-charge condition therefore reduces, with the bulk potential taken as the reference, to
\begin{equation}
	\left.\frac{d\phi^0}{dr}\right|_{r=1}=-\sigma,\qquad \phi^0(r)\to0\quad\text{as}\quad r\to\infty .
	\label{eq:PB_bc}
\end{equation}
The constant interior equilibrium potential is then determined from the potential-continuity condition in Eq.~\eqref{eq:8}, giving
\begin{equation}
	\phi_d^0=\phi^0(1).
	\label{eq:phi_d_equilibrium}
\end{equation}
The equilibrium variables $\phi^0(r)$ and $n_i^0(r)$ are subsequently retained as known coefficients in the $O(\Lambda)$ electrokinetic problem.
\subsection{First-order electrokinetic perturbation}
Since the equilibrium state is spherically symmetric and the imposed electric field is uniform along the $z$-axis, the leading-order disturbance is axisymmetric about this axis and has dipolar angular dependence. This is consistent with the far-field condition $\phi\sim-\Lambda r\cos\theta$ as $r\to \infty$. The scalar perturbations may therefore be expanded in the first spherical harmonic as \cite{baygents1991electrophoresis,ohshima1984electrokinetic,jayaraman2019unusual}
\begin{equation}
	\phi^{(1)}(r,\theta)=-\Psi(r)\cos\theta,\qquad \phi_d^{(1)}(r,\theta)=-\Psi_d(r)\cos\theta,\qquad \mu_i^{(1)}(r,\theta)=-z_i\Phi_i(r)\cos\theta .
	\label{eq:first_harmonic_scalar}
\end{equation}
Here \(\Psi(r)\), \(\Psi_d(r)\) and \(\Phi_i(r)\) are the radial amplitudes of the exterior electric-potential perturbation, interior electric-potential perturbation and electrochemical-potential perturbation, respectively. Expanding \(\mu_i=\mu_i^{\rm ref}+z_i\phi+\ln n_i\) to \(O(\Lambda)\) gives
\begin{equation}
	\mu_i^{(1)}=z_i\phi^{(1)}+\frac{n_i^{(1)}}{n_i^0}.
	\label{eq:first_order_mu_relation}
\end{equation}
Using Eq.~\eqref{eq:first_harmonic_scalar}, the concentration perturbation is
\begin{equation}
	n_i^{(1)}(r,\theta)=z_i n_i^0(r)[\Psi(r)-\Phi_i(r)]\cos\theta .
	\label{eq:first_order_concentration}
\end{equation}
Equation \eqref{eq:first_order_concentration}  shows that the ionic concentration induced by the imposed electric field can be determined once the electric-potential and electrochemical-potential perturbations are known.
The velocity field is also of order $O(\Lambda)$. Owing to the same axisymmetry and the dipolar angular dependence imposed by the far-field electric field, the first-order incompressible velocity field may be expressed in terms of a single scalar radial function \cite{landau1987fluid,jayaraman2019unusual,mahapatra2022electrophoresis}. Thus, the exterior perturbation velocity is written as
\begin{equation}
	\bm{u}^{(1)}(r,\theta)=-\frac{2H(r)}{r}\cos\theta\,\bm{e}_r+\frac{1}{r}\frac{d}{dr}\{rH(r)\}\sin\theta\,\bm{e}_{\theta}.
	\label{eq:u_expansion}
\end{equation}
Similarly, the velocity field inside the porous droplet is represented as
\begin{equation}
	\bm{u}_d^{(1)}(r,\theta)=-\frac{2H_d(r)}{r}\cos\theta\,\bm{e}_r+\frac{1}{r}\frac{d}{dr}\{rH_d(r)\}\sin\theta\,\bm{e}_{\theta}.
	\label{eq:ud_expansion}
\end{equation}
The forms \eqref{eq:u_expansion} and \eqref{eq:ud_expansion} automatically satisfy the incompressibility conditions \eqref{eq:2b} and \eqref{eq:3b}, respectively. For compactness, we introduce the radial operator associated with the first spherical harmonic as $L f=d^2f/dr^2+(2/r)df/dr-2f/r^2$.
\subsubsection{\bf Perturbed electric potential outside the droplet}
The first-order electrostatic equation follows by substituting Eqs.~\eqref{eq:first_harmonic_scalar} and \eqref{eq:first_order_concentration} into the exterior Poisson equation \eqref{eq:6}. Collecting the terms proportional to \(\Lambda\cos\theta\) gives
\begin{equation}
	L\Psi=\frac{(\kappa a)^2}{2}\sum_{i=1}^{N}z_i^2 n_i^0(r)[\Psi(r)-\Phi_i(r)],\qquad r>1 .
	\label{eq:perturbed_Poisson}
\end{equation}
This equation describes the distortion of the electric potential induced by the imposed electric field. Its coupling with $\Phi_i$ represents the relaxation of the ionic cloud around the porous droplet. Inside the ion-free porous droplet, the perturbed electric potential satisfies Laplace's equation \eqref{eq:7}, and hence
\begin{equation}
	L\Psi_d=0,\qquad r<1 .
	\label{eq:perturbed_Laplace_inside}
\end{equation}
The solution of Eq.\eqref{eq:perturbed_Laplace_inside} that remains bounded at the origin is $\Psi_d=A_d r$, where $A_d$ is determined by the interfacial electric conditions. The \(O(\Lambda)\) part of the potential-continuity condition in Eq.~\eqref{eq:8} gives
\begin{equation}
	\Psi(1)=\Psi_d(1).
	\label{eq:perturbed_phi_continuity}
\end{equation}
Since the surface charge density is prescribed and fixed, its first-order perturbation vanishes. Hence the \(O(\Lambda)\) part of the normal electric-displacement condition in Eq.~\eqref{eq:8} gives
\begin{equation}
	\Psi'(1)-\varepsilon_r\Psi_d'(1)=0.
	\label{eq:perturbed_displacement}
\end{equation}
Using \(\Psi_d=A_d r\), Eqs.~\eqref{eq:perturbed_phi_continuity} and \eqref{eq:perturbed_displacement} reduce to
\begin{equation}
	\Psi'(1)=\varepsilon_r\Psi(1).
	\label{eq:perturbed_electric_BC_reduced}
\end{equation}
Finally, the $O(\Lambda)$ part of the far-field condition \(\phi\sim-\Lambda r\cos\theta\) in Eq.~\eqref{eq12}, together with the expansion \eqref{eq:first_harmonic_scalar}, gives
\begin{equation}
	\Psi(r)\sim r\qquad\text{as}\qquad r\to\infty .
	\label{eq:perturbed_phi_farfield}
\end{equation}
\subsubsection{\bf Perturbed ion-transport equations}
Expanding the ion-flux conservation law \eqref{eq5} to $O(\Lambda)$, and using the incompressibility of $\bm{u}^{(1)}$, gives
\begin{equation}
	\nabla\cdot\left(n_i^0\nabla\mu_i^{(1)}\right)=Pe_i\,\bm{u}^{(1)}\cdot\nabla n_i^0 .
	\label{eq:ion_transport_first_order_general}
\end{equation}
Substitution of \(\mu_i^{(1)}=-z_i\Phi_i(r)\cos\theta\) and Eq.~\eqref{eq:u_expansion} into Eq.~\eqref{eq:ion_transport_first_order_general} gives the radial equation
\begin{equation}
	L\Phi_i=-\frac{d}{dr}\left(\ln n_i^0\right)\frac{d\Phi_i}{dr}+\frac{2Pe_i}{z_i}\frac{d}{dr}\left(\ln n_i^0\right)\frac{H}{r},\qquad r>1 .
	\label{eq:perturbed_Phi_equation}
\end{equation}
The first term on the right-hand side arises from the equilibrium concentration gradient in the electric double layer, whereas the second term represents the convective distortion of the equilibrium ionic cloud by the electrophoretic flow. Further, the ion-impermeability condition \eqref{eq6} at the droplet surface, together with the normal-velocity condition $v(1)=0$, gives
\begin{equation}
	\Phi_i'(1)=0.
	\label{eq:Phi_surface_BC}
\end{equation}
Far from the droplet, the ionic concentrations return to their bulk values. Since \(n_i^{(1)}\to0\) and \(\Psi(r)\sim r\), Eq.~\eqref{eq:first_order_concentration} gives
\begin{equation}
	\Phi_i(r)\sim r\qquad\text{as}\qquad r\to\infty .
	\label{eq:Phi_farfield_BC}
\end{equation}
\subsubsection{\bf Perturbed hydrodynamics outside the droplet}
The exterior momentum equation \eqref{eq:3a} contains the electrical body force exerted by the charged diffuse layer. At $O(1)$, this body force is spherically symmetric and is balanced by the equilibrium pressure. At $O(\Lambda)$, the imposed field distorts the double layer and generates a solenoidal electrohydrodynamic forcing. Taking the curl of the first-order momentum equation eliminates pressure and gives
\begin{equation}
	\nabla\times\nabla\times\nabla\times\bm{u}^{(1)}=\frac{(\kappa a)^2}{2}\sum_{i=1}^{N}\nabla\mu_i^{(1)}\times\nabla n_i^0 .
	\label{eq:hydro_curl_general}
\end{equation}
Using \(\mu_i^{(1)}=-z_i\Phi_i\cos\theta\) and the velocity representation \eqref{eq:u_expansion}, Eq.~\eqref{eq:hydro_curl_general} reduces to
\begin{equation}
	L(LH)=\frac{(\kappa a)^2}{2r}\sum_{i=1}^{N}z_i\frac{dn_i^0}{dr}\Phi_i,\qquad r>1 .
	\label{eq:H_equation}
\end{equation}
Since the forcing on the right-hand side of \eqref{eq:H_equation} decays away from the electric double layer, the exterior flow far from the droplet satisfies the homogeneous equation $L(LH)=0$, as $r\rightarrow\infty$, which has the general solution of the form
\begin{equation}
	H(r)=A r^3+B r+C+\frac{D}{r^2}.
	\label{eq:H_farfield_general}
\end{equation}
The term $A r^3$ gives a velocity that grows algebraically at infinity and is therefore inadmissible. The term $B r$ represents the imposed uniform flow in the droplet-fixed frame. Indeed, using the velocity representation \eqref{eq:u_expansion}, the far-field velocity associated with $H\sim Br$ is $v^{(1)}\sim -2B\cos\theta$ and $u^{(1)}\sim 2B\sin\theta$. In the reference frame attached to the droplet, the first-order far-field fluid velocity must approach $-\mu\bm{e}_z$, as given in Eq.~\eqref{eq12}, where $\mu=U_E/\Lambda$ is the dimensionless electrophoretic mobility. Since $-\mu\Lambda\bm{e}_z=-\mu\Lambda\cos\theta\bm{e}_r+\mu\Lambda\sin\theta\bm{e}_\theta$, comparison gives $2B=\mu$. Further, the constant term $C$ in \eqref{eq:H_farfield_general} produces a velocity disturbance of order $1/r$, which corresponds to a Stokeslet contribution. Since the electrophoretic droplet is freely suspended, the total hydrodynamic force on it must vanish; hence the Stokeslet contribution is absent and $C=0$. The remaining term $D/r^2$ gives a faster-decaying disturbance and is admissible. Therefore, the physically acceptable far-field behaviour is
\begin{equation}
	H(r)=\frac{\mu}{2}r+O(r^{-2}),\qquad r\rightarrow\infty .
	\label{eq:H_farfield}
\end{equation}
Rather than prescribing the unknown mobility $\mu$ in advance, the far-field behaviour \eqref{eq:H_farfield} is imposed through derivative conditions. From \eqref{eq:H_farfield}, one obtains
\begin{equation}
	H''(r)\rightarrow0,\qquad r^2\frac{d}{dr}\left(\frac{H}{r}\right)\rightarrow0,\qquad r\rightarrow\infty .
	\label{eq:H_farfield_conditions}
\end{equation}
The first condition eliminates the algebraically growing mode, while the second condition removes the Stokeslet mode and therefore enforces the force-free nature of electrophoretic motion. After solving the boundary-value problem for $H(r)$, the electrophoretic mobility is extracted from the coefficient of the uniform far-field flow as
\begin{equation}
	\mu=\lim_{r\to\infty}\frac{2H(r)}{r}.
\end{equation}
\subsubsection{\bf Perturbed hydrodynamics inside the porous droplet}
The key distinction between a clean liquid droplet and the present porous droplet appears in the internal hydrodynamic equation. The $O(\Lambda)$ velocity field inside the droplet satisfies the Brinkman equation 
\begin{equation}
	\eta_r\nabla\times\nabla\times\bm{u}_d^{(1)}+\nabla p_d^{(1)}+\frac{\eta_r}{Da}\bm{u}_d^{(1)}=0,\qquad \nabla\cdot\bm{u}_d^{(1)}=0.
	\label{eq:internal_Brinkman}
\end{equation}
Taking the curl of Eq.~\eqref{eq:internal_Brinkman} eliminates the pressure perturbation. Substitution of the velocity representation \eqref{eq:ud_expansion} then gives
\begin{equation}
	L(LH_d)-\frac{1}{Da}LH_d=0,\qquad r<1.
	\label{eq:Hd_ODE}
\end{equation}
The internal velocity field must remain bounded at the droplet centre. Therefore,
\begin{equation}
	H_d(r)\quad \text{is regular as}\quad r\rightarrow0.
	\label{eq:Hd_regular}
\end{equation}
\subsubsection{\bf Interfacial hydrodynamic boundary conditions}
The hydrodynamic boundary conditions at \(r=1\) follow from Eq.~\eqref{eq:4}. The normal-velocity condition gives
\begin{equation}
	H(1)=0,\qquad H_d(1)=0.
	\label{eq:H_no_penetration}
\end{equation}
Continuity of tangential velocity gives
\begin{equation}
	H'(1)=H_d'(1).
	\label{eq:H_tangential_velocity}
\end{equation}
The tangential stress balance is obtained from Eq.~\eqref{eq:5}. Using Eq.~\eqref{eq:u_expansion} and the no-penetration condition, the hydrodynamic shear terms at the interface reduce to
\begin{equation}
	\left(\frac{\partial u}{\partial r}-\frac{u}{r}\right)_{r=1^+}=\Lambda H''(1)\sin\theta,\qquad \left(\frac{\partial u_d}{\partial r}-\frac{u_d}{r}\right)_{r=1^-}=\Lambda H_d''(1)\sin\theta .
	\label{eq:tangential_stress_components}
\end{equation}
The \(O(\Lambda)\) tangential Maxwell traction generated by the fixed surface charge is
\begin{equation}
	-\sigma\left.\frac{\partial\phi}{\partial\theta}\right|_{r=1}=-\Lambda\sigma\Psi(1)\sin\theta .
	\label{eq:Maxwell_traction_first_order}
\end{equation}
Therefore, the first-order tangential stress balance becomes
\begin{equation}
	H''(1)-\eta_r H_d''(1)=\sigma\Psi(1).
	\label{eq:H_tangential_stress}
\end{equation}
This condition is where Brinkman-screened internal hydrodynamics meets the tangential electric traction, exterior hydrodynamic shear and internal hydrodynamic shear and together determine the interfacial motion and its transmission into internal circulation.

Combining the preceding results, the perturbation problem created by the imposed electric field is therefore reduced to the coupled radial system \eqref{eq:perturbed_Poisson}, \eqref{eq:perturbed_Phi_equation}, \eqref{eq:H_equation} and \eqref{eq:Hd_ODE}, subject to the electric boundary conditions \eqref{eq:perturbed_electric_BC_reduced} and \eqref{eq:perturbed_phi_farfield}, the ion-transport boundary conditions \eqref{eq:Phi_surface_BC} and \eqref{eq:Phi_farfield_BC}, the far-field hydrodynamic condition \eqref{eq:H_farfield_conditions}, the interfacial hydrodynamic conditions \eqref{eq:H_no_penetration}, \eqref{eq:H_tangential_velocity} and \eqref{eq:H_tangential_stress}, and the regularity condition at droplet's centre \eqref{eq:Hd_regular}. The solution of this boundary-value problem determines the electric-potential perturbation, ionic relaxation, exterior electrohydrodynamic flow, Brinkman-screened internal circulation and electrophoretic mobility.
\section{\bf Analytical solution under the Debye--H\"uckel approximation}
\label{sec:DH_solution}
We now derive an analytical expression for the electrophoretic mobility in the Debye--H\"uckel regime. The analysis is restricted to a symmetric monovalent electrolyte, with $z_+=1$ and $z_-=-1$, and to small equilibrium surface potential such that $|\zeta|\ll1$. Under this approximation, the equilibrium ion concentrations are linearized as \(n_\pm^0\simeq1\mp\phi^0\), and Eq.~\eqref{eq:equilibrium_PB} reduces to the linearized Poisson--Boltzmann equation.

For compactness, we denote $q=\kappa a$. Solving the linearized form of Eq.~\eqref{eq:equilibrium_PB} with the boundary conditions \eqref{eq:PB_bc} gives
\begin{equation}
	\phi^0(r)=\frac{\sigma}{1+q}\frac{\exp[-q(r-1)]}{r},
	\qquad 
	\zeta=\phi^0(1)=\frac{\sigma}{1+q}.
	\label{eq:DH_phi0}
\end{equation}
Here $\zeta$ is the equilibrium surface potential. In the Debye--H\"uckel limit, terms nonlinear in $\zeta$ are neglected throughout the analytical calculation.

\subsection{\bf Analytic expressions for perturbed electrochemical and electric potentials}
In the Debye--H\"uckel regime, the perturbed electrochemical potential \eqref{eq:perturbed_Phi_equation} reduces to $L\Phi_i=0$ subject to the boundary conditions \eqref{eq:Phi_surface_BC} and \eqref{eq:Phi_farfield_BC}. Hence, for both ionic species,
\begin{equation}
	\Phi_i(r)=\Phi(r)=r+\frac{1}{2r^2}.
	\label{eq:DH_Phi_solution}
\end{equation}
The perturbed electric-potential equation \eqref{eq:perturbed_Poisson} becomes $L\Psi=q^2(\Psi-\Phi)$, where $\Phi$ is given by Eq.~\eqref{eq:DH_Phi_solution}. The solution satisfying $\Psi(r)\sim r$ as $r\rightarrow\infty$ can be written as
\begin{equation}
	\Psi(r)=r+\frac{1}{2r^2}+C_\varepsilon f(r),
	\qquad 
	f(r)=\frac{\exp(-qr)}{q^2}\left(\frac{1}{r^2}+\frac{q}{r}\right).
	\label{eq:DH_Psi_solution}
\end{equation}
The constant $C_\varepsilon$ is determined from the interfacial electric condition \eqref{eq:perturbed_electric_BC_reduced}, yielding
\begin{equation}
	C_\varepsilon=\frac{3\varepsilon_r}{2[f'(1)-\varepsilon_r f(1)]}
	=-\frac{3\varepsilon_r q^2\exp(q)}{2\{q^2+(2+\varepsilon_r)q+(2+\varepsilon_r)\}}.
	\label{eq:Cepsilon}
\end{equation}
Therefore, the value of the perturbed potential at the interface is
\begin{equation}
	\Psi(1)=\frac{3}{2}\left[1-\frac{\varepsilon_r(q+1)}{q^2+(2+\varepsilon_r)q+(2+\varepsilon_r)}\right].
	\label{eq:Psi_surface_DH}
\end{equation}
For a non-polarizable droplet, $\varepsilon_r=0$, and Eq.~\eqref{eq:Psi_surface_DH} reduces to $\Psi(1)=3/2$. Further, the Maxwell traction can be obtained by substituting $\Psi(1)$ into the Eq.\eqref{eq:Maxwell_traction_first_order} as 
\begin{equation}
	-\sigma\left.\frac{\partial\phi}{\partial\theta}\right|_{r=1} (=-\Lambda\sigma\Psi(1)\sin\theta)=-\frac{3}{2}\zeta(q+1) \left[1-\frac{\varepsilon_r(q+1)}{q^2+(2+\varepsilon_r)q+(2+\varepsilon_r)}\right]\Lambda\sin\theta.
	\label{eq:Maxwell_traction_expression}
\end{equation}
Here we have used the Debye--H\"uckel relation $\sigma=\zeta(q+1)$. The bracketed factor is positive for $q>0$ and $\varepsilon_r\geq0$, since $1-\varepsilon_r(q+1)[q^2+(2+\varepsilon_r)q+(2+\varepsilon_r)]^{-1}=(q^2+2q+2)[q^2+(2+\varepsilon_r)q+(2+\varepsilon_r)]^{-1}>0$. Thus, the signed coefficient of the tangential Maxwell traction is proportional to, and opposite in sign to, the zeta potential \(\zeta\). The bracketed factor quantifies the reduction of this traction by dielectric polarization of the droplet interior; it is unity for a non-polarizable droplet $(\varepsilon_r=0)$ and tends to zero in the highly polarizable limit $\varepsilon_r\to\infty$.
\subsection{\bf Solution for Brinkman-screened internal flow and interfacial resistance}\label{Brink_hydro}
The internal flow field is governed by Eq.\eqref{eq:Hd_ODE}, which is repeated here as
\begin{equation}
	L(LH_d)-\frac{1}{Da}LH_d=0.
	\label{eq:DH_Hd_eq}
\end{equation}
Introducing the Brinkman screening parameter $\chi=Da^{-1/2}$, Eq.~\eqref{eq:DH_Hd_eq} may be written as $(L-\chi^{2})LH_d=0$. The general solution of this equation can be expressed as 
\begin{equation}
	H_d(r)=C_1 r+\frac{C_2}{r^2}+C_3 i_1(\chi r)+C_4 k_1(\chi r),
	\label{eq:Hd_general_solution}
\end{equation}
where $i_1$ and $k_1$ are the modified spherical Bessel functions of order one of the first and second kinds, respectively.

The regularity condition \eqref{eq:Hd_regular} implies $C_{2}=0=C_{4}$. The no-penetration condition \eqref{eq:H_no_penetration} gives \(C_1=-C_3 i_1(\chi)\), and the tangential-velocity continuity condition \eqref{eq:H_tangential_velocity} gives \(C_3=H'(1)/[\chi i_1'(\chi)-i_1(\chi)]\), where the prime on \(i_1\) denotes differentiation with respect to its argument. Thus, the solution for $H_d(r)$, which is regular at the centre of the droplet $r=0$, can be written as
\begin{equation}
	H_d(r)=H'(1)\left[\frac{i_1(\chi r)-r i_1(\chi)}{\chi i_1'(\chi)-i_1(\chi)}\right].
	\label{eq:Hd_solution_Bessel}
\end{equation}
Differentiating Eq.~\eqref{eq:Hd_solution_Bessel} twice with respect to $r$ and evaluating at the interface gives the interfacial velocity--shear relation
\begin{equation}
	H_d''(1)=\mathcal{R}(\chi)H'(1),
	\label{eq:Hd_second_relation}
\end{equation}
where, after using the explicit hyperbolic representation of $i_1$, the function $\mathcal{R}(\chi)$ is obtained as
\begin{equation}
	\mathcal{R}(\chi)=
	\frac{\chi(\chi^2+6)\cosh\chi-3(\chi^2+2)\sinh\chi}
	{(\chi^2+3)\sinh\chi-3\chi\cosh\chi}.
	\label{eq:R_chi}
\end{equation}
Eq.\eqref{eq:Hd_second_relation} shows that \(\mathcal{R}(\chi)\) is the Brinkman-screened hydrodynamic resistance of the porous droplet interior. It depends only on the permeability through \(\chi=Da^{-1/2}\) and is independent of the electrostatic parameters. Substitution of Eq.~\eqref{eq:Hd_second_relation} into the tangential stress condition \eqref{eq:H_tangential_stress} gives the effective interfacial stress balance
\begin{equation}
	H''(1)-\eta_r\mathcal{R}(\chi)H'(1)=\sigma\Psi(1).
	\label{eq:DH_effective_stress_BC}
\end{equation}
In the classical clean-droplet problem, the internal Stokes flow contributes the factor $3\eta_r$ to the interfacial stress balance. In the present porous-droplet problem, this factor is $\eta_r\mathcal{R}(\chi)$. Equation~\eqref{eq:DH_effective_stress_BC} thus shows that the porous interior enters the electrophoretic problem by replacing the clean-droplet internal Stokes resistance \(3\eta_r\) with the Brinkman-screened resistance \(\eta_r\mathcal{R}(\chi)\). Hence, the effect of internal permeability is incorporated through a single resistance function.

In the clean-droplet limit, \(Da\to\infty\) or \(\chi\to0\), one obtains \(\mathcal{R}(\chi)\to3\), and Eq.~\eqref{eq:DH_effective_stress_BC} recovers the classical tangential stress condition for a clean liquid droplet. In contrast, as \(Da\to0\), \(\mathcal{R}(\chi)\) becomes large, indicating strong suppression of internal circulation by the porous matrix. Thus, $\mathcal{R}(\chi)$ connects the clean-droplet limit to the rigid-particle-like limit.
\subsection{\bf Explicit mobility expression for a polarizable porous droplet}
Under the Debye--H\"uckel approximation, the exterior hydrodynamic equation \eqref{eq:H_equation} reduces to
\begin{equation}
	L(LH)=G(r),
	\label{eq:DH_H_eq}
\end{equation}
where, using Eqs.~\eqref{eq:DH_phi0} and \eqref{eq:DH_Phi_solution},
\begin{equation}
	G(r)=q^2\zeta\left(\frac{1+qr}{r^2}\right)\left(1+\frac{1}{2r^3}\right)\exp[-q(r-1)].
	\label{eq:G_DH}
\end{equation}
The boundary conditions for $H$ are $H(1)=0$, the effective stress condition \eqref{eq:DH_effective_stress_BC}, and the far-field conditions corresponding to a force-free electrophoretic motion. Solving Eq.~\eqref{eq:DH_H_eq} using these boundary conditions gives the electrophoretic mobility in the integral form
\begin{equation}
	\mu=\frac{1}{9}\int_1^\infty
	\left[
	2\left(\frac{\eta_r\mathcal{R}(\chi)+3}{\eta_r\mathcal{R}(\chi)+2}\right)r^3
	-3r^2
	+\frac{\eta_r\mathcal{R}(\chi)}{\eta_r\mathcal{R}(\chi)+2}
	\right]G(r)\,dr
	-\frac{2\sigma\Psi(1)}{3[\eta_r\mathcal{R}(\chi)+2]}.
	\label{eq:mobility_integral_DH}
\end{equation}
The first term represents the contribution from the electrical body force distributed within the electric double layer, whereas the second term arises from the tangential Maxwell stress acting directly at the charged interface.

The integral in Eq.~\eqref{eq:mobility_integral_DH} can be evaluated explicitly. Defining the exponential integral of order $n$ as $E_n(q)=\int_1^\infty t^{-n}\exp(-qt)\,dt$, the electrophoretic mobility becomes
\begin{equation}
	\mu=
	\zeta\left[
	\frac{q+\eta_r\mathcal{R}(\chi)+1}{\eta_r\mathcal{R}(\chi)+2}
	+2e^qE_5(q)
	-\frac{5\eta_r\mathcal{R}(\chi)}{\eta_r\mathcal{R}(\chi)+2}e^qE_7(q)
	-\frac{2(q+1)}{3[\eta_r\mathcal{R}(\chi)+2]}\Psi(1)
	\right],
	\label{eq:mobility_DH_general}
\end{equation}
where $\mathcal{R}(\chi)$ and $\Psi(1)$ are respectively given by Eq.~\eqref{eq:R_chi} and Eq.~\eqref{eq:Psi_surface_DH}. Equation~\eqref{eq:mobility_DH_general} is the Debye--H\"uckel mobility of a porous liquid droplet with fixed surface charge. It shows explicitly how the electrostatic response, through $\Psi(1)$, and the internal hydrodynamic resistance, through $\eta_r\mathcal{R}(Da^{-1/2})$, jointly determine the electrophoretic mobility. This expression may also be used by the experimentalist to estimate the $\zeta$-potential from measured mobility data once the permeability, viscosity ratio, dielectric ratio and double-layer thickness are known. It is worth noting that the Debye--H\"uckel mobility of a porous droplet is obtained from the corresponding clean-droplet mobility by replacing the classical internal Stokes resistance $3\eta_r$ with the Brinkman-screened resistance $\eta_r\mathcal{R}(Da^{-1/2})$.

For a non-polarizable porous droplet, $\varepsilon_r=0$ and hence $\Psi(1)=3/2$. Equation~\eqref{eq:mobility_DH_general} then reduces to
\begin{equation}
	\mu=
	\zeta\left[
	\frac{\eta_r\mathcal{R}(\chi)}{\eta_r\mathcal{R}(\chi)+2}
	+2e^qE_5(q)
	-\frac{5\eta_r\mathcal{R}(\chi)}{\eta_r\mathcal{R}(\chi)+2}e^qE_7(q)
	\right].
	\label{eq:mobility_DH_nonpolarizable}
\end{equation}
\subsection{\bf Decomposition of mobility into rigid-particle and internal-circulation contributions}
For a non-polarizable porous droplet, Eq.~\eqref{eq:mobility_DH_nonpolarizable} can be rearranged as
\begin{equation}
	\mu=\mu^{R}
	+\frac{2\zeta}{\eta_r\mathcal{R}(\chi)+2}
	\left[5e^qE_7(q)-1\right],
	\label{eq:mobility_decomposition}
\end{equation}
where $\mu^{R}=\zeta\left[1+2e^qE_5(q)-5e^qE_7(q)\right]$ is the Debye--H\"uckel mobility of a rigid spherical particle \cite{ohshima1983approximate}. The second term in Eq.~\eqref{eq:mobility_decomposition} represents the correction caused by interfacial mobility and internal circulation. This correction is controlled by the factor $[\eta_r\mathcal{R}(\chi)+2]^{-1}$ and therefore vanishes either when $Da\to0$ or when $\eta_r\to\infty$. Thus, the Brinkman resistance suppresses the liquid-droplet contribution and continuously drives the electrophoretic response toward the rigid-particle limit.

The role of the Darcy number is explicit through $\chi=Da^{-1/2}$ and $\mathcal{R}(\chi)$. In the highly permeable limit $Da\to\infty$, $\mathcal{R}(\chi)\to3$, and Eq.~\eqref{eq:mobility_DH_nonpolarizable} recovers the classical clean-droplet result. In the opposite limit $Da\to0$, $\mathcal{R}(\chi)\to\infty$, and the mobility approaches the rigid-particle mobility, $\mu=\zeta\left[1+2e^qE_5(q)-5e^qE_7(q)\right]$. This shows that permeability controls the strength of the liquid-droplet correction to the rigid-particle mobility.

Since $5e^qE_7(q)-1<0$ for $q\ge0$, the internal-circulation correction has a sign opposite to $\zeta$ and therefore reduces the magnitude of the fixed-charge mobility relative to the rigid-particle value.
\subsection{\bf Interfacial velocity, internal circulation and their correlation with mobility}
\label{sec:interface_circulation_DH}

The decomposition in Eq.~\eqref{eq:mobility_decomposition} suggests that the deviation of a porous droplet from the rigid-particle mobility is caused by finite interfacial motion and the associated internal circulation. We now make this connection explicit. From the velocity representation in Eq.~\eqref{eq:u_expansion}, the tangential velocity at the droplet surface is $u_s(\theta)=u(1,\theta)=\Lambda H'(1)\sin\theta$. Hence the maximum interfacial velocity occurs at the equator and is given by $u_s^{m}=\Lambda H'(1)$. From the exterior hydrodynamic solution of Eq.~\eqref{eq:DH_H_eq}, the derivative of $H$ at the droplet surface can be expressed as
\begin{equation}
	H'(1)=\frac{1}{3[\eta_r\mathcal{R}(\chi)+2]}
	\int_1^\infty (r^3-1)G(r)\,dr
	-\frac{\sigma\Psi(1)}{\eta_r\mathcal{R}(\chi)+2}.
	\label{eq:Hprime_integral_DH}
\end{equation}
The first term arises from the electrical body force distributed within the electric double layer, whereas the second term arises from the tangential Maxwell stress acting at the charged interface. Since the denominator in $H'(1)$ is positive, Eq.\eqref{eq:Hprime_integral_DH} shows that the direction of interfacial motion is determined entirely by the signed competition between the diffuse-layer hydrodynamic forcing and the tangential Maxwell traction. Substituting the Debye--H\"uckel expression for $G(r)$ given in (\ref{eq:G_DH}) and using $\sigma=(1+q)\zeta$, Eq.~\eqref{eq:Hprime_integral_DH} gives
\begin{equation}
	H'(1)=\frac{3\zeta}{2[\eta_r\mathcal{R}(\chi)+2]}
	\left[
	q-1+10e^qE_7(q)-\frac{2(q+1)}{3}\Psi(1)
	\right].
	\label{eq:Hprime_general_DH}
\end{equation}
For a non-polarizable porous droplet, $\varepsilon_r=0$ and $\Psi(1)=3/2$, so that
\begin{equation}
	H'(1)=\frac{3\zeta}{\eta_r\mathcal{R}(\chi)+2}
	\left[5e^qE_7(q)-1\right].
	\label{eq:Hprime_nonpolarizable_DH}
\end{equation}
Since $5e^qE_7(q)-1<0$ for $q\ge0$, the interfacial velocity for a non-polarizable porous droplet has a sign opposite to $\zeta$. Thus the direction of the equatorial interfacial motion is fixed by the sign of the surface charge, while its magnitude is reduced by the Brinkman resistance through the factor $[\eta_r\mathcal{R}(\chi)+2]^{-1}$.

Subtracting the rigid-particle mobility from Eq.~\eqref{eq:mobility_DH_general} and using Eq.~\eqref{eq:Hprime_general_DH}, we obtain
\begin{equation}
	\mu-\mu^R=\frac{2}{3}H'(1)=\frac{2}{3}\frac{u_s^m}{\Lambda}.
	\label{eq:mobility_surface_velocity_relation}
\end{equation}
Thus, the departure from rigid-particle electrophoresis is directly proportional to the first interfacial-velocity mode. When the interfacial velocity vanishes, the porous droplet electrophoretically behaves as a rigid particle.

We next quantify the strength of the internal circulation. The azimuthal vorticity associated with the internal velocity field is
\begin{equation}
	\omega_{d,\varphi}
	=
	\left[
	\frac{\partial u_d}{\partial r}
	+\frac{u_d}{r}
	-\frac{1}{r}\frac{\partial v_d}{\partial\theta}
	\right]
	=
	\Lambda L H_d(r)\sin\theta .
	\label{eq:internal_vorticity}
\end{equation}
Accordingly, we define a signed circulation measure over a meridional half-plane as
\begin{equation}
	\Omega=\int_0^1\int_0^\pi
	\omega_{d,\varphi}r\,d\theta\,dr =\Lambda \int_0^1\int_0^\pi
	rL H_d(r)\sin\theta \,d\theta\,dr.
	\label{eq:Omega_definition}
\end{equation}
Using the Brinkman solution \eqref{eq:Hd_solution_Bessel}, we obtain
\begin{equation}
	\Omega=\Lambda\,\mathcal{C}(\chi)H'(1),
	\label{eq:Omega_Hprime_relation}
\end{equation}
where the circulation-response function is
\begin{equation}
	\mathcal{C}(\chi)=
	\frac{2\int_0^\chi s\,i_1(s)\,ds}
	{\chi i_1'(\chi)-i_1(\chi)} .
	\label{eq:C_chi}
\end{equation}
The function $\mathcal{C}(\chi)$ is positive and measures how effectively a given interfacial velocity generates circulation inside the porous droplet. In the clean-droplet limit, $\chi\to0$, one obtains $\mathcal{C}(\chi)\to10/3$, so that $\Omega=(10/3)\Lambda H'(1)$, which is the classical result for a clean liquid droplet \cite{majhi2025electrophoresis}. In the strongly resistive limit, $Da\to0$, the interfacial velocity itself vanishes as a consequence of the large Brinkman resistance, and hence the internal circulation is suppressed.

Combining Eqs.~\eqref{eq:mobility_surface_velocity_relation} and \eqref{eq:Omega_Hprime_relation}, we obtain a direct relation between electrophoretic mobility and internal circulation:
\begin{equation}
	\mu=\mu^R+\frac{2}{3\mathcal{C}(\chi)}\frac{\Omega}{\Lambda}.
	\label{eq:mobility_Omega_relation}
\end{equation}
Equation~\eqref{eq:mobility_Omega_relation} shows that the departure from rigid-particle electrophoresis is controlled by the strength and direction of the Brinkman-screened internal circulation. When $\Omega=0$, the porous droplet has no internal circulation and its mobility becomes identical to that of a rigid particle. Conversely, finite circulation produces a liquid-droplet correction to the rigid-particle mobility.

For a charged droplet with $\zeta\neq0$, the reversal of interfacial motion and internal circulation occurs when $H'(1)=0$. From Eq.~\eqref{eq:Hprime_general_DH}, the reversal condition is
\begin{equation}
	q-1+10e^qE_7(q)-\frac{2(q+1)}{3}\Psi(1)=0.
	\label{eq:circulation_reversal_condition}
\end{equation}
Since $\Omega$ is proportional to $H'(1)$, the same condition determines the reversal of internal circulation. For a non-polarizable droplet, Eq.~\eqref{eq:circulation_reversal_condition} reduces to $5e^qE_7(q)-1=0$, which has no finite positive root because $5e^qE_7(q)-1<0$ for $q\ge0$. Thus, in the non-polarizable fixed-charge case, the circulation direction does not reverse with $q$ or $Da$; it changes only with the sign of $\zeta$. For a polarizable porous droplet, however, $\Psi(1)$ depends on $\varepsilon_r$, and the reversal condition is modified by dielectric polarization. Depending on $\varepsilon_r$, this condition may admit a finite positive root. The Darcy number does not set the reversal condition directly; rather, it controls the magnitude of the circulation through the resistance factor $[\eta_r\mathcal{R}(\chi)+2]^{-1}$.
\subsection{\bf Limiting cases of the Debye--H\"uckel mobility}
\label{sec:DH_limiting_cases}
We now examine several limiting cases of Eq.~\eqref{eq:mobility_DH_general}. These limits are important because they validate the analytical expression and clarify the physical role of the Darcy number.

First, consider the clean-droplet limit. When the Darcy number becomes very large, $Da\to\infty$, the internal porous resistance vanishes and $\chi\to0$. In this limit, $\mathcal{R}(\chi)\to3$. Substituting into Eq.~\eqref{eq:mobility_DH_general} gives
\begin{equation}
	\mu=
	\zeta\left[
	\frac{q+3\eta_r+1}{3\eta_r+2}
	+2e^qE_5(q)
	-\frac{15\eta_r}{3\eta_r+2}e^qE_7(q)
	-\frac{2(q+1)}{3(3\eta_r+2)}\Psi(1)
	\right].
	\label{eq:mobility_clean_limit_general}
\end{equation}
This is precisely the Debye--H\"uckel mobility of a clean liquid droplet with fixed surface charge \cite{mahapatra2022electrophoresis}. Therefore, the present porous-droplet theory continuously recovers the classical liquid-droplet result when the internal permeability becomes sufficiently large.

For a non-polarizable droplet, $\varepsilon_r=0$, and hence $\Psi(1)=3/2$. Equation~\eqref{eq:mobility_clean_limit_general} then reduces to
\begin{equation}
	\mu=
	\zeta\left[
	\frac{3\eta_r}{3\eta_r+2}
	+2e^qE_5(q)
	-\frac{15\eta_r}{3\eta_r+2}e^qE_7(q)
	\right],
	\qquad Da\to\infty,\quad \varepsilon_r=0.
	\label{eq:mobility_clean_nonpolarizable}
\end{equation}
This is the standard Debye--H\"uckel result for a non-polarizable liquid droplet with immobile surface charge and is the same expression as derived by Booth \cite{booth1951cataphoresis} for a non-conducting droplet.

In the opposite limit, $Da\to0$, the porous resistance becomes dominant and $\chi\to\infty$. The resistance function then grows without bound, $\mathcal{R}(\chi)\to\infty$. Consequently, Eq.~\eqref{eq:mobility_DH_general} becomes
\begin{equation}
	\mu=\zeta\left[1+2e^qE_5(q)-5e^qE_7(q)\right],
	\qquad Da\to0.
	\label{eq:mobility_rigid_limit}
\end{equation}
This is the Debye--H\"uckel electrophoretic mobility of a rigid spherical particle \cite{ohshima1983approximate}. Therefore, a strongly resistive porous droplet behaves hydrodynamically like a rigid particle because the internal circulation is suppressed by the Brinkman drag. The same rigid-particle limit is also obtained when the droplet viscosity is very large. Indeed, for fixed $Da$, taking $\eta_r\to\infty$ gives $\eta_r\mathcal{R}(\chi)\to\infty$, and Eq.~\eqref{eq:mobility_DH_general} again reduces to Eq.~\eqref{eq:mobility_rigid_limit}. Thus, both vanishing permeability and very large internal viscosity suppress the internal circulation and lead to rigid-particle-like electrophoresis.

Next, we examine the H\"uckel limit, $q=\kappa a\to0$. Since $e^qE_5(q)\to1/4$ and $e^qE_7(q)\to1/6$ as $q\to0$, Eq.~\eqref{eq:mobility_DH_general} gives
\begin{equation}
	\mu=
	\zeta\left[
	\frac{2\{\eta_r\mathcal{R}(\chi)+3\}}{3\{\eta_r\mathcal{R}(\chi)+2\}}
	-\frac{2}{(\varepsilon_r+2)\{\eta_r\mathcal{R}(\chi)+2\}}
	\right],
	\qquad q\to0.
	\label{eq:mobility_Huckel_general}
\end{equation}
For a non-polarizable porous droplet, $\varepsilon_r=0$, this simplifies to
\begin{equation}
	\mu=
	\zeta\,
	\frac{2\eta_r\mathcal{R}(\chi)+3}
	{3\{\eta_r\mathcal{R}(\chi)+2\}},
	\qquad q\to0,\quad \varepsilon_r=0.
	\label{eq:mobility_Huckel_nonpolarizable}
\end{equation}
If the droplet is also clean, so that $Da\to\infty$ and $\mathcal{R}(\chi)\to3$, Eq.~\eqref{eq:mobility_Huckel_nonpolarizable} becomes
\begin{equation}
	\mu=
	\zeta\,\frac{2\eta_r+1}{3\eta_r+2},
	\qquad q\to0,\quad Da\to\infty,\quad \varepsilon_r=0.
	\label{eq:Huckel_clean_drop}
\end{equation}
On the other hand, when $Da\to0$ or $\eta_r\to\infty$, Eq.~\eqref{eq:mobility_Huckel_nonpolarizable} gives $\mu=(2/3)\zeta$, $q\to0$, which is the classical H\"uckel mobility of a rigid sphere.

Finally, we consider the thin-double-layer limit, $q=\kappa a\to\infty$. In this limit, $e^qE_5(q)$ and $e^qE_7(q)$ are both $O(q^{-1})$ and therefore vanish at leading order. The surface value of the perturbed potential gives a finite contribution through the Maxwell-stress term, and Eq.~\eqref{eq:mobility_DH_general} reduces to
\begin{equation}
	\mu=
	\zeta\,
	\frac{\eta_r\mathcal{R}(\chi)+\varepsilon_r}
	{\eta_r\mathcal{R}(\chi)+2},
	\qquad q\to\infty.
	\label{eq:mobility_Smol_general}
\end{equation}
For a non-polarizable porous droplet, this becomes
\begin{equation}
	\mu=
	\zeta\,
	\frac{\eta_r\mathcal{R}(\chi)}
	{\eta_r\mathcal{R}(\chi)+2},
	\qquad q\to\infty,\quad \varepsilon_r=0.
	\label{eq:mobility_Smol_nonpolarizable}
\end{equation}
In the clean-droplet limit, Eq.~\eqref{eq:mobility_Smol_nonpolarizable} further reduces to
\begin{equation}
	\mu=
	\zeta\,\frac{3\eta_r}{3\eta_r+2},
	\qquad q\to\infty,\quad Da\to\infty,\quad \varepsilon_r=0.
	\label{eq:Smol_clean_drop}
\end{equation}
In contrast, for $Da\to0$ or $\eta_r\to\infty$, Eq.~\eqref{eq:mobility_Smol_nonpolarizable} gives $\mu=\zeta$, $q\to\infty$,
which is the classical Smoluchowski mobility \cite{von1921elektrische} of a rigid sphere.

These limiting cases confirm that the present Debye--H\"uckel expression has the correct behaviour in all relevant regimes. The Darcy number provides a continuous hydrodynamic bridge between a clean liquid droplet and a rigid particle: large $Da$ permits interfacial motion and internal circulation, whereas small $Da$ suppresses the internal flow and recovers the rigid-particle mobility.
\section{Numerical implementation}\label{numerical_method}
The coupled radial boundary-value problem obtained from the weak-field perturbation formulation in section \ref{sec:perturbation} is solved numerically using COMSOL Multiphysics. Since the equilibrium variables and perturbation amplitudes depend only on the radial coordinate, a one-dimensional computational domain $1\leq r\leq R_\infty$ is used, where $R_\infty=100$ denotes the truncated far-field boundary. Further increase in $R_\infty$ does not produce any change in the computed solution.

The nonlinear equilibrium Poisson--Boltzmann equation \eqref{eq:equilibrium_PB} is solved using the ``Coefficient Form PDE'' interface. The resulting equilibrium potential $\phi^0(r)$ is then used to evaluate the equilibrium ion concentrations $n_i^0(r)$ from the Boltzmann distribution \eqref{eq:equilibrium_boltzmann}. These equilibrium fields are then used as known coefficients in the subsequent first-order perturbation problem.

The first-order perturbed equations \eqref{eq:perturbed_Poisson}, \eqref{eq:perturbed_Phi_equation} and \eqref{eq:H_equation} are solved using the interface ``Coefficient Form PDE'' for the five exterior radial unknowns $\Phi_+(r)$, $\Phi_-(r)$, $\Psi(r)$, $Z(r)$ and $H(r)$. Here, $Z=LH$ is introduced to reduce the fourth-order hydrodynamic equation to a coupled system of second-order equations. The boundary conditions at $r=1$ and $r=R_\infty$ are imposed according to the weak-field formulation described in section~\ref{sec:perturbation}. At the finite outer boundary $r=R_\infty$, the far-field conditions are imposed in their asymptotic form.

The internal Brinkman hydrodynamic field is not discretized numerically, because the internal equation \eqref{eq:Hd_ODE} admits a regular analytical solution, as derived in subsection~\ref{Brink_hydro}. This solution is valid for any $Da>0$ without imposing any asymptotic restriction on the Darcy number, and gives the interfacial relation $H_d''(1)=\mathcal{R}(\chi)H'(1)$, where $\chi=Da^{-1/2}$. Hence the internal flow enters the exterior numerical problem only through the tangential stress condition \eqref{eq:H_tangential_stress}. Since \(Z=LH\) and \(H(1)=0\), one has \(Z(1)=H''(1)+2H'(1)\). Combining this identity with \eqref{eq:H_tangential_stress} gives the effective boundary condition
\begin{equation}
	Z(1)=\left[2+\eta_r \mathcal{R}(\chi)\right]H'(1)+\sigma\Psi(1).
	\label{eq:effective_Z_BC}
\end{equation}
This treatment avoids numerical discretization of the region \(0\leq r<1\), while retaining the full Brinkman-screened hydrodynamic resistance of the porous interior over the range of Darcy numbers considered in the present study.

Each radial equation is implemented in COMSOL using the stationary ``Coefficient Form PDE'' interface, whose general form is
\begin{equation}
	\nabla\cdot\left(-c\nabla X-\boldsymbol{\alpha}X+\boldsymbol{\gamma}\right)+\boldsymbol{\beta}\cdot\nabla X+aX=f,
	\label{eq:comsol_coefficient_form}
\end{equation}
where $X$ denotes the dependent variable, $c$ is the diffusion coefficient, $\boldsymbol{\alpha}$ and $\boldsymbol{\beta}$ are conservative and convective coefficient vectors, $\boldsymbol{\gamma}$ is the conservative source term, $a$ is the absorption coefficient and $f$ is the source term. In the present one-dimensional radial implementation, Eq.~\eqref{eq:comsol_coefficient_form} reduces to
\begin{equation}
	\frac{d}{dr}\left(-c\frac{dX}{dr}-\alpha X+\gamma\right)+\beta\frac{dX}{dr}+aX=f.
\end{equation}
The coefficients $c$, $\alpha$, $\gamma$, $\beta$, $a$ and $f$ are specified separately for each dependent variable. For example, the equilibrium Poisson--Boltzmann equation \eqref{eq:equilibrium_PB} for a symmetric monovalent electrolyte is implemented by choosing $X=\phi^0$, $c=-1$, $\alpha=0$, $\gamma=0$, $\beta=2/r$, $a=0$ and $f=(\kappa a)^2\sinh \phi^0$.

A non-uniform mesh is used to resolve the electric double layer accurately. The finest spacing is prescribed near the droplet surface, where the electrostatic and hydrodynamic fields vary most rapidly. Specifically, $\Delta r=0.01/(\kappa a)$ is used in the near-interfacial region $1\leq r\leq 1+2/(\kappa a)$, corresponding to approximately the first two Debye lengths outside the droplet. Away from the electrical double layer, the mesh is gradually coarsened to $\Delta r=0.01$. The dependent variables were approximated using fifth-order Lagrange elements, and the resulting nonlinear algebraic system is solved using a fully coupled stationary Newton solver.

After convergence, the electrophoretic mobility is obtained from the far-field coefficient of the hydrodynamic solution as $\mu={2H(R_\infty)}/{R_\infty}$. The interfacial velocity mode $H'(1)$, hydrodynamic shear stress and tangential Maxwell traction are evaluated directly from the converged radial solution using Eqs.~\eqref{eq:tangential_stress_components} and \eqref{eq:Maxwell_traction_first_order}.
\section{\bf Results and discussion}\label{Discussions}
The numerical results are obtained using parameter values representative of aqueous electrolyte systems at room temperature. In particular, the thermal potential is taken as $\phi_0=k_BT/e=25.8~{\rm mV}$, the viscosity of the external electrolyte as $\eta=0.89\times10^{-3}~{\rm Pa\,s}$, and the elementary charge as $e=1.602\times10^{-19}~{\rm C}$. The droplet radius is fixed at $a=50~{\rm nm}$. The suspending electrolyte is taken to be KCl, for which the ionic diffusivities are $D_{\mathrm{K}^+}=1.96\times10^{-9}~{\rm m^2\,s^{-1}}$ and $D_{\mathrm{Cl}^-}=2.03\times10^{-9}~{\rm m^2\,s^{-1}}$. The dielectric permittivity of the electrolyte is $\varepsilon_e=78.54\varepsilon_0$, where $\varepsilon_0$ is the permittivity of free space.

The principal dimensionless parameters varied in the computations are the Darcy number $Da$, the viscosity ratio $\eta_r$, the permittivity ratio $\varepsilon_r$, the surface charge density $\sigma^*$, and the Debye-layer parameter $q=\kappa a$. The viscosity ratio is varied over $10^{-2}\leq\eta_r\leq10^{3}$, covering low-viscosity droplets such as bubbles through to highly viscous droplets approaching the rigid-particle limit. The permittivity ratio is varied over $0\leq\varepsilon_r\leq10^{3}$; the limit $\varepsilon_r=0$ represents a non-polarizable droplet, whereas $\varepsilon_r=10^{3}$ represents a highly polarizable dielectric droplet. The Darcy number is varied over $10^{-5}\leq Da\leq10$, which spans strongly Brinkman-screened droplets to highly permeable droplets. The surface charge density is varied up to a magnitude of $24~{\rm mC\,m^{-2}}$, and the Debye-layer parameter is varied over $1\leq q\leq 125$.
\subsection{Validation of the numerical results with Debye--H\"uckel analytical solution}
\begin{figure}[H]
	\centering
	\begin{minipage}[h]{0.32\textwidth}
		\centering
		\begin{overpic}[width=\linewidth]{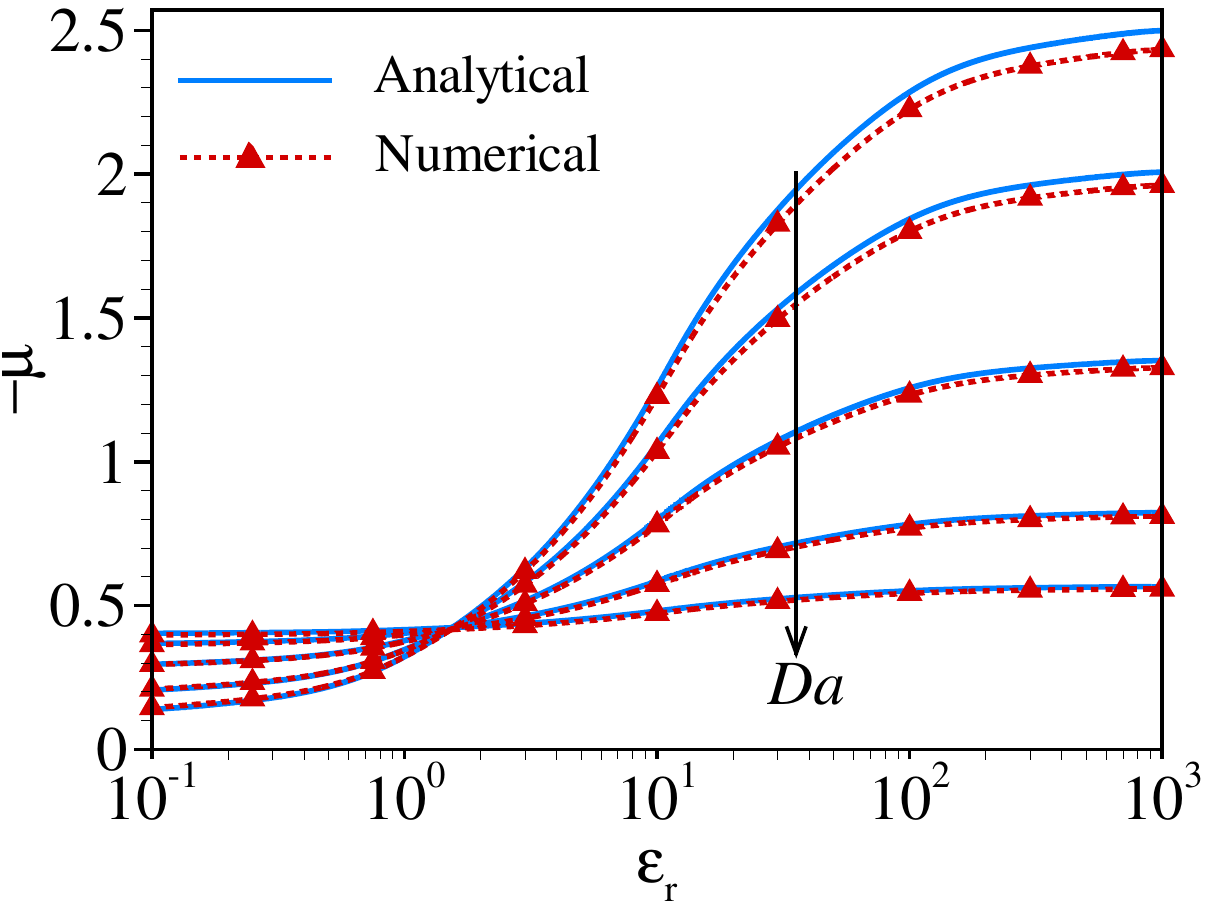}
			\put(2,80){(a)}
		\end{overpic}
	\end{minipage}
	\begin{minipage}[h]{0.32\textwidth}
		\centering
		\begin{overpic}[width=\linewidth]{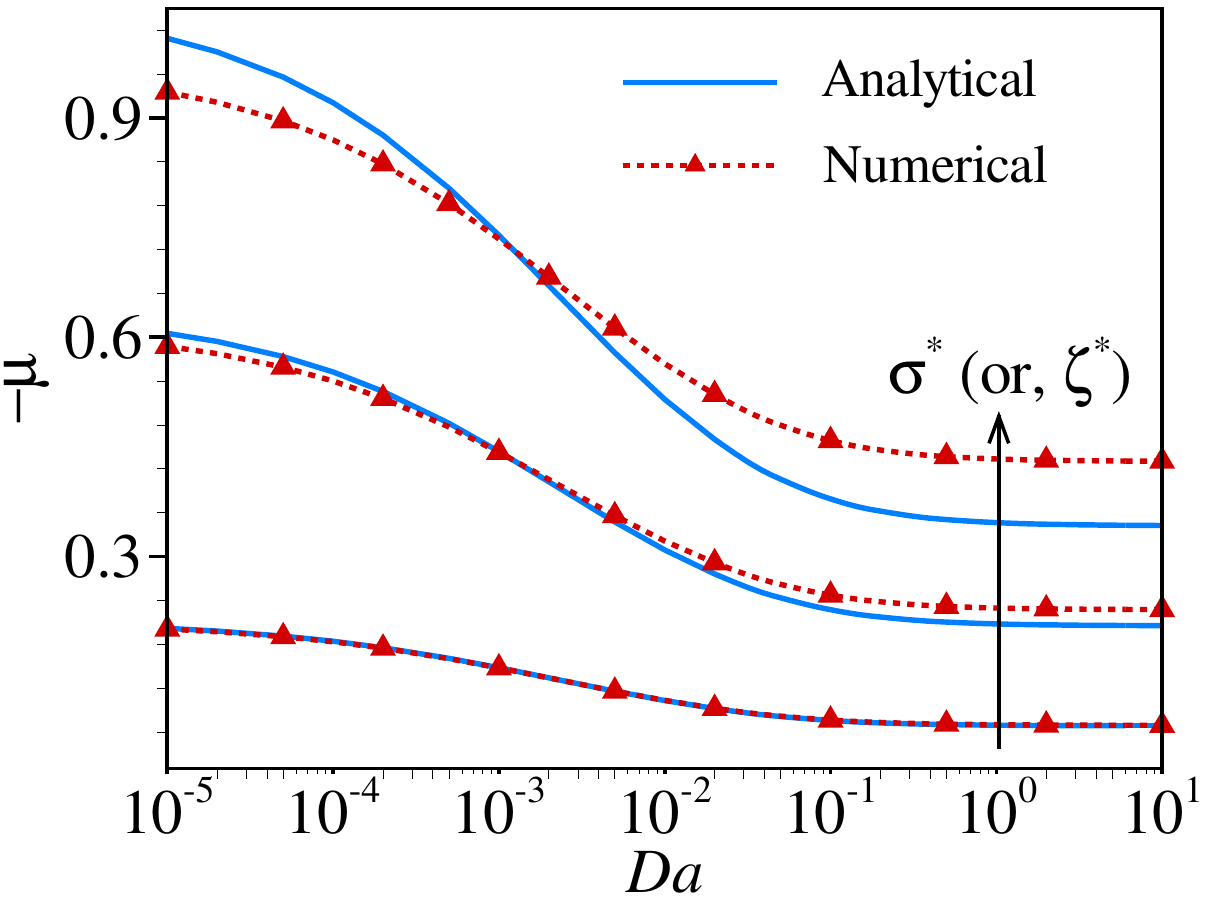}
			\put(2,80){(b)}
		\end{overpic}
	\end{minipage}
	\begin{minipage}[h]{0.32\textwidth}
		\centering
		\begin{overpic}[width=\linewidth]{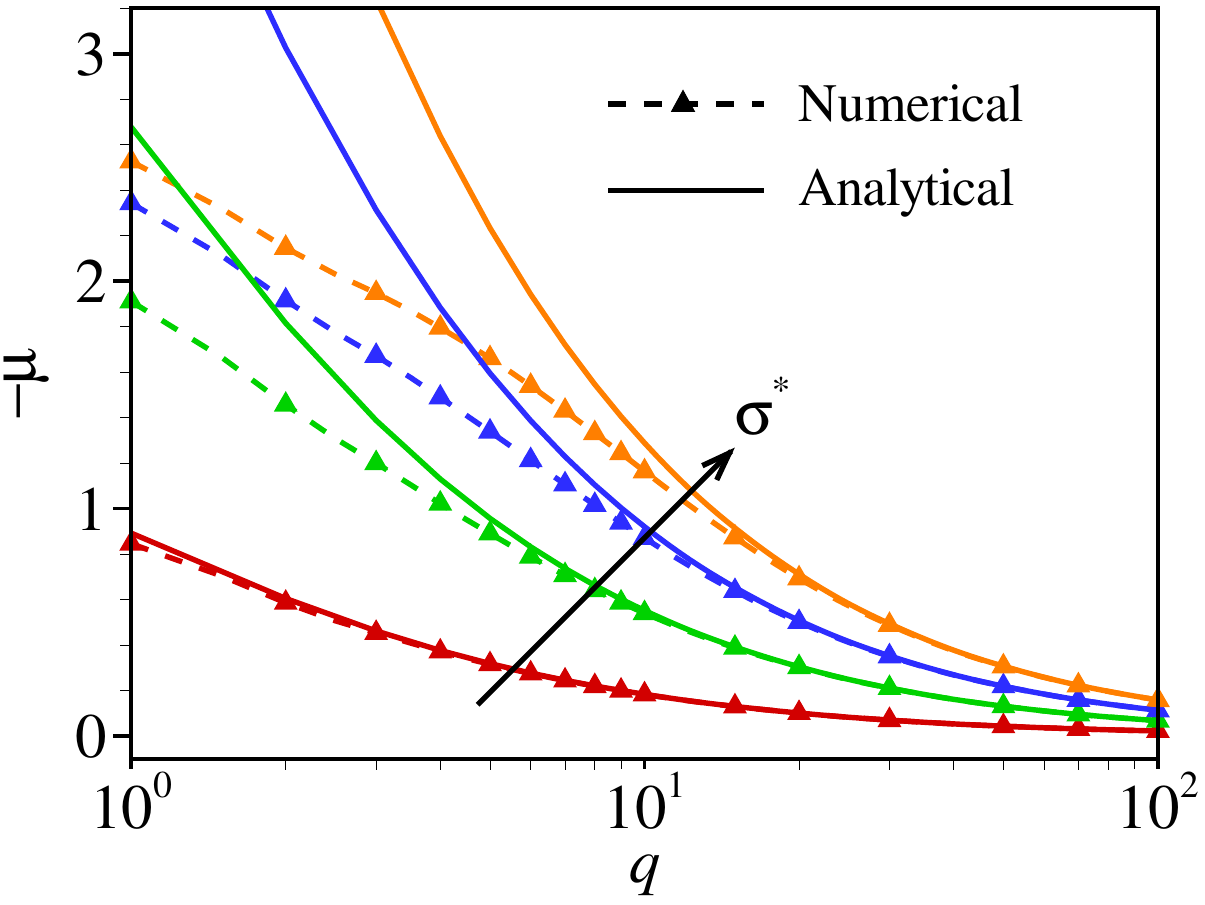}
			\put(2,80){(c)}
		\end{overpic}
	\end{minipage}
	\caption{Comparison between the numerical solution and the analytical solution for the magnitude of the electrophoretic mobility, $-\mu$, as a function of (a) $\varepsilon_{r}$ for different values of the Darcy number, $Da=10^{-5},10^{-3},10^{-2},10^{-1},1$ at $\sigma^{*}=-2~{\rm mC\,m^{-2}}$ and $q=10$; (b) $Da$ for different values of $\sigma^{*}=-1~{\rm mC\,m^{-2}}, -3~{\rm mC\,m^{-2}}, -5~{\rm mC\,m^{-2}}$ at $\varepsilon_{r}=0.1$ and $q=10$ and (c) $q$ for different values of $\sigma^{*}=-1~{\rm mC\,m^{-2}}, -3~{\rm mC\,m^{-2}}, -5~{\rm mC\,m^{-2}}, -7~{\rm mC\,m^{-2}}$ at $\varepsilon_{r}=0.1$ and $Da=10^{-4}$. The viscosity ratio is $\eta_r=0.1$ in all panels.}
	\label{fig:AA}
\end{figure}
We first validate the numerical solution by comparing it with the Debye--H\"uckel analytical mobility given in Eq.~\eqref{eq:mobility_DH_general}. Figure~\ref{fig:AA}(a) shows the variation of the mobility with the permittivity ratio $\varepsilon_r$ for several values of the Darcy number $Da$, at a small surface charge density. The numerical results are in excellent agreement with the analytical prediction over the entire range of $\varepsilon_r$ and $Da$ considered. This agreement is expected because Eq.~\eqref{eq:mobility_DH_general} is valid for arbitrary values of $Da$, $\varepsilon_r$ and $q=\kappa a$, provided the equilibrium interfacial potential remains small compared with the thermal potential. The range of validity of the Debye--H\"uckel expression is examined more directly in Fig.~\ref{fig:AA}(b), where the mobility is plotted as a function of $\sigma^*$ at fixed $q$. The analytical and numerical solutions agree closely for small $|\sigma^*|$, but a clear deviation appears as the magnitude of the surface charge is increased. This departure is the breakdown of the low-potential approximation used in deriving Eq.~\eqref{eq:mobility_DH_general}. For example, when $\sigma^*=-5~{\rm mC\,m^{-2}}$, the corresponding equilibrium surface potential becomes $-31~{\rm mV}$, larger in magnitude than the thermal voltage $\phi_0=25.8~\mathrm{mV}$, so nonlinear double-layer polarization can no longer be neglected. Figure~\ref{fig:AA}(c) further illustrates this point by comparing the analytical and numerical mobilities over a broad range of $q$. At low surface charge, the Debye--H\"uckel expression remains accurate for all values of $q$ considered. At larger $|\sigma^*|$, however, the discrepancy becomes pronounced in the small-$q$ regime. This is because, for fixed surface charge density, a thicker double layer produces a larger equilibrium surface potential. Consequently, nonlinear double-layer effects become stronger when $q$ is small and $|\sigma^*|$ is large. As $q$ increases, the same surface charge is screened over a thinner layer and the equilibrium surface potential decreases. The solution therefore re-enters the Debye--H\"uckel regime, and the analytical and numerical results again approach one another. Thus, Fig.~\ref{fig:AA} confirms two points. First, the numerical method correctly reproduces the closed-form Debye--H\"uckel mobility in its range of validity. Secondly, the deviations observed at larger surface charge or smaller $q$ are physically meaningful, reflecting nonlinear electric-double-layer polarization beyond the low-potential approximation. 
\subsection{Permeability-controlled mobility: effect of Darcy number and dielectric polarizability}
\begin{figure}[H]
	\centering
	\begin{minipage}[h]{0.32\textwidth}
		\centering
		\begin{overpic}[width=\linewidth]{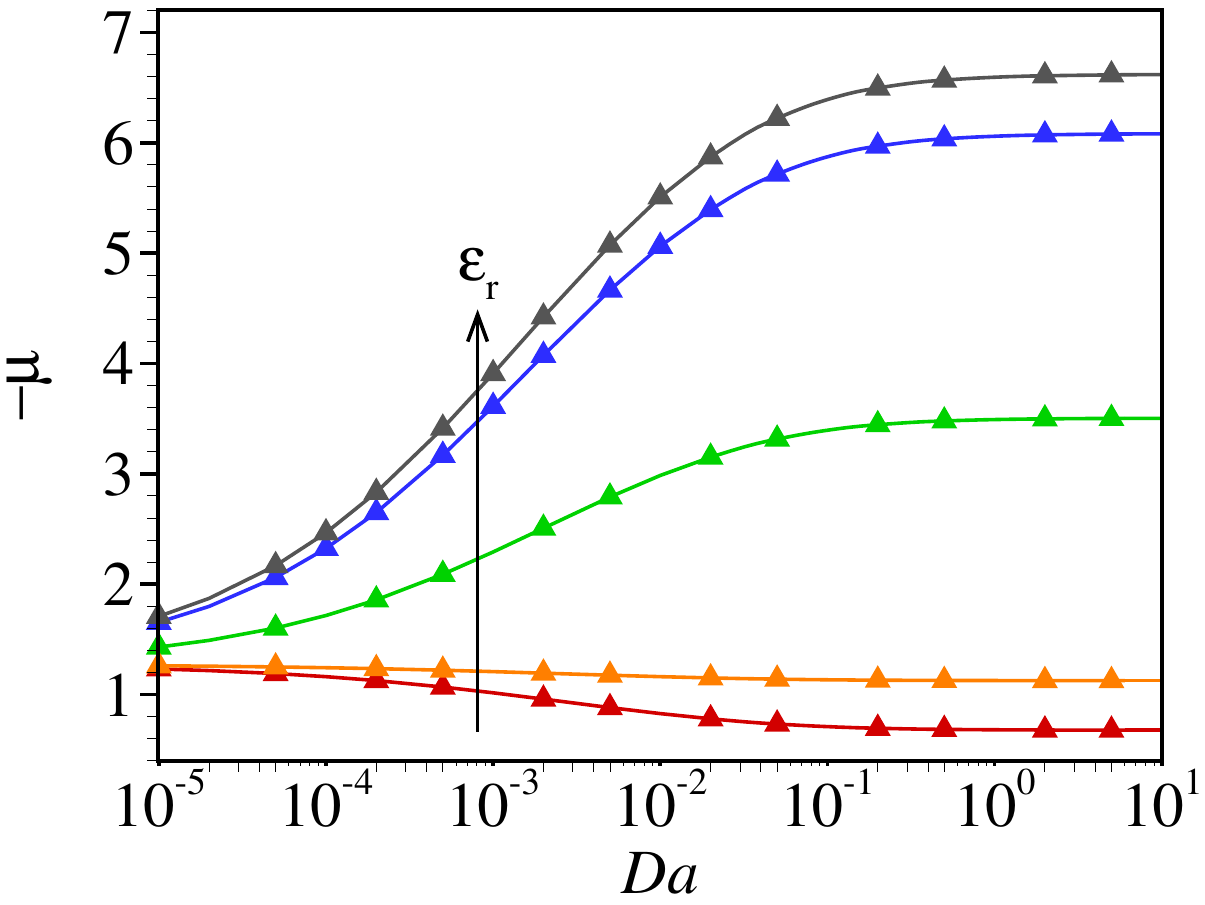}
			\put(2,77){(a)}
		\end{overpic}
	\end{minipage}
	\begin{minipage}[h]{0.32\textwidth}
		\centering
		\begin{overpic}[width=\linewidth]{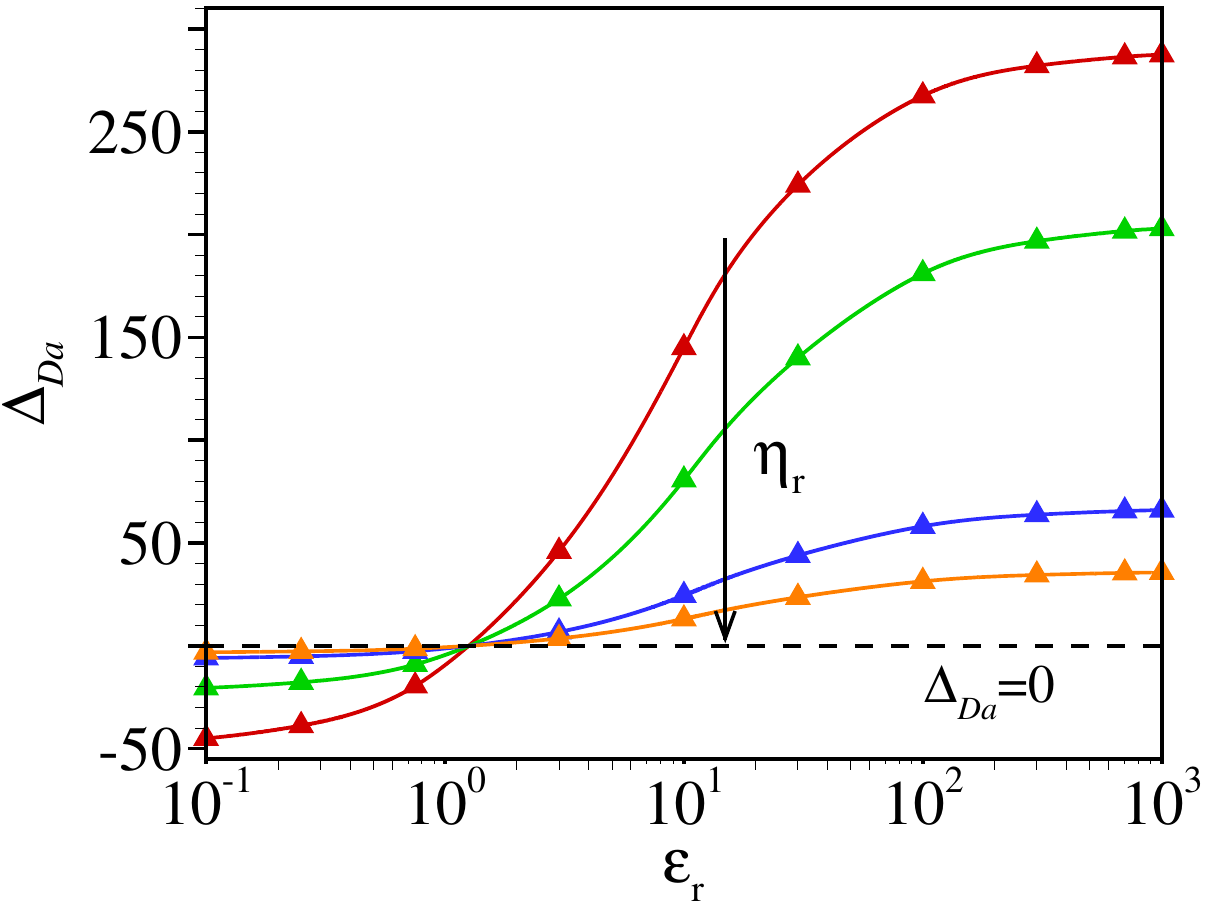}
			\put(2,77){(b)}
		\end{overpic}
	\end{minipage}
	\begin{minipage}[h]{0.32\textwidth}
		\centering
		\begin{overpic}[width=\linewidth]{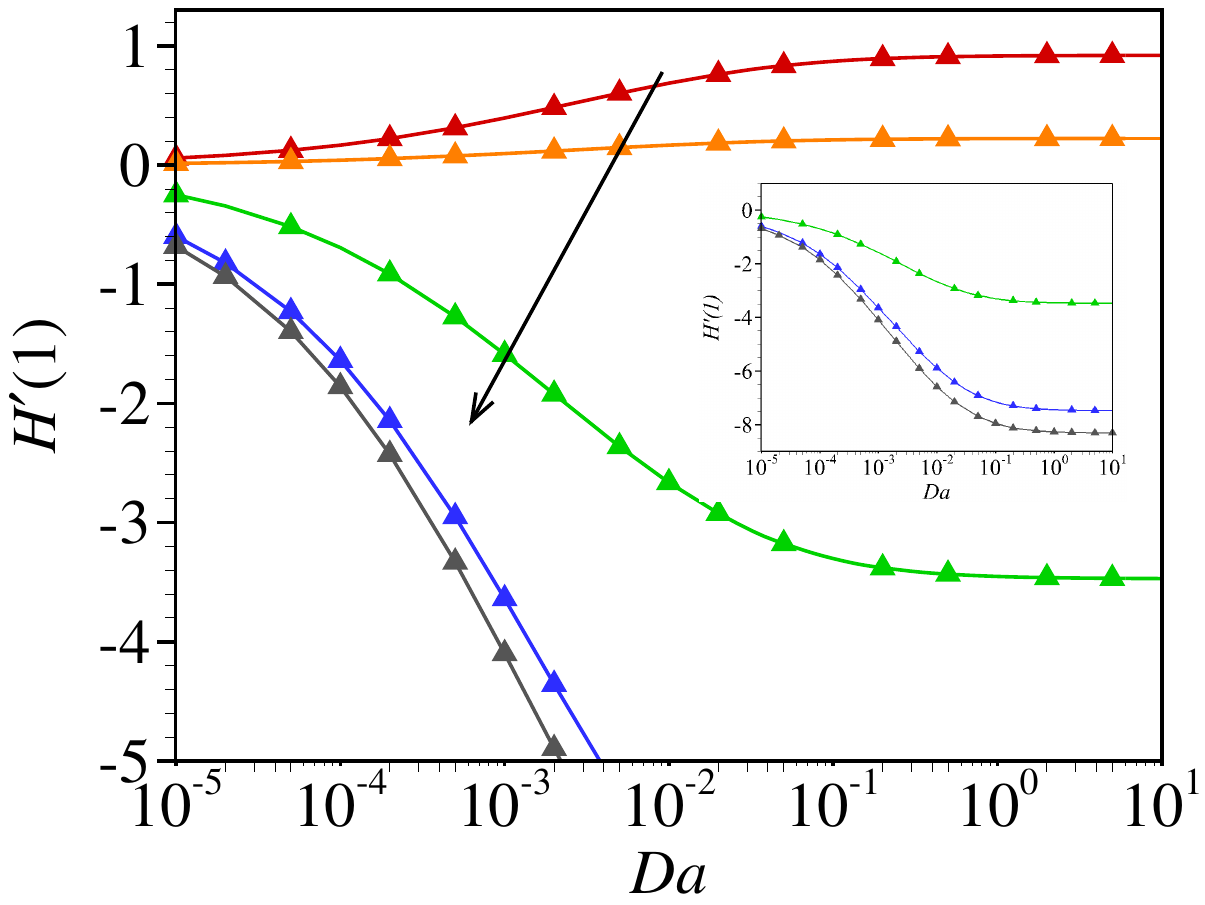}
			\put(2,77){(c)}
		\end{overpic}
	\end{minipage}
	\caption{
		Variation of (a) the electrophoretic mobility $\mu$, with the Darcy number $Da$ for different dielectric permittivity ratio $e_r=0.1,1,10,100,1000$ at $\eta_r=0.1$; (b) the relative Darcy response, $\Delta_{Da}=100\{[-\mu]_{Da=10}-[-\mu]_{Da=10^{-5}}\}/[-\mu]_{Da=10^{-5}}$, as a function of $e_r$ for different viscosity ratio $\eta_r=0.1,1,5,10$; and (c) the interfacial velocity measure $H'(1)$ as a function of $Da$ for different dielectric permittivity ratio $e_r=0.1,1,10,100,1000$ at $\eta_r=0.1$. The other parameters for panels (a)--(c) are $\kappa a=10$ and $\sigma^*=-7~\mathrm{mC\,m^{-2}}$.}
	\label{fig:BB}
\end{figure}
We now examine how the permeability of the porous interior modifies the electrophoresis of the droplet. A purely hydrodynamic intuition might suggest that increasing $Da$ should always promote internal circulation and thereby enhance droplet mobility. However, Fig.\ref{fig:BB}(a) shows that the effect of permeability is not universal, but depends strongly on the dielectric polarizability of the droplet. For weakly polarizable droplets, represented by small values of $\varepsilon_r$, the magnitude of the mobility decreases as $Da$ increases. In contrast, for highly polarizable droplets, the opposite trend is observed. In this case, increasing $Da$ enhances the magnitude of the electrophoretic mobility. This behaviour is summarized in Fig.~\ref{fig:BB}(b), where we plot the relative Darcy response
\begin{equation}
	\Delta_{Da}
	=
	100\frac{[-\mu]_{Da=10}-[-\mu]_{Da=10^{-5}}}{[-\mu]_{Da=10^{-5}}}.
	\label{eq:Delta_Da_results}
\end{equation}
A negative value of $\Delta_{Da}$ indicates that increasing permeability reduces the mobility magnitude, whereas a positive value indicates that increasing permeability enhances it. The zero crossing of $\Delta_{Da}$ therefore identifies a transition between permeability-suppressed and permeability-enhanced electrophoresis. For low $\varepsilon_r$, $\Delta_{Da}<0$, while for sufficiently large $\varepsilon_r$, $\Delta_{Da}>0$. The transition is weakened as the viscosity ratio increases, because a larger $\eta_r$ increases the effective internal hydrodynamic resistance and thereby reduces the sensitivity of the mobility to the Darcy number.

The mechanism behind this transition is revealed by Fig.~\ref{fig:BB}(c), which shows the interfacial-velocity measure $H'(1)$. From the velocity representation, $H'(1)$ determines the tangential velocity at the droplet surface, and hence measures the strength and direction of the interfacial motion created by the imposed electric field. Since the internal circulation strength is proportional to $H'(1)$, the sign reversal of $H'(1)$ also represents a reversal of the Brinkman-screened circulation inside the porous droplet. The analytical decomposition derived earlier shows that the departure from rigid-particle mobility is controlled by this interfacial mode, namely
\begin{equation}
	\mu-\mu^R=\frac{2}{3}H'(1).
	\label{eq:mu_Hprime_discussion}
\end{equation}
Since the surface charge considered here is negative, the rigid-particle contribution to the mobility is negative. Therefore, when $H'(1)>0$, the circulation correction in Eq.~\eqref{eq:mu_Hprime_discussion} is positive and opposes the negative rigid-particle contribution; as a result, the magnitude $-\mu$ decreases as permeability increases. This is precisely the behaviour observed for weakly polarizable droplets. Conversely, when $H'(1)<0$, the circulation correction has the same sign as the rigid-particle mobility and reinforces the electrophoretic motion; consequently, $-\mu$ increases with $Da$, as observed for highly polarizable droplets.

The reversal of the Darcy-number response therefore originates from a reversal of the interfacial velocity. For weakly polarizable droplets, the electrohydrodynamically induced interfacial motion generates an internal circulation whose contribution to the mobility opposes the rigid-particle electrophoretic contribution, thereby reducing the mobility magnitude relative to the rigid-particle limit. For highly polarizable droplets, dielectric polarization modifies the perturbed electric field at the interface and reverses the direction of the interfacial motion. Once this reversal occurs, increasing permeability strengthens a circulation mode that assists, rather than opposes, electrophoretic migration. The Darcy number therefore acts as a hydrodynamic amplifier of the interfacial motion determined by the electrostatic boundary-value problem.

It is important to emphasize that $Da$ does not directly modify the electric double layer. The double-layer structure is controlled by $q$, $\sigma^*$ and the electrostatic boundary condition involving $\varepsilon_r$. The role of $Da$ is hydrodynamic. It controls how the interfacial motion generated by the electric and hydrodynamic stresses at the interface is transmitted into the Brinkman-screened interior. 
\subsection{Surface-charge and Debye-layer effects on the permeability response}
\begin{figure}[H]
	\centering
	\begin{minipage}[h]{0.3\textwidth}
		\centering
		\begin{overpic}[width=\linewidth]{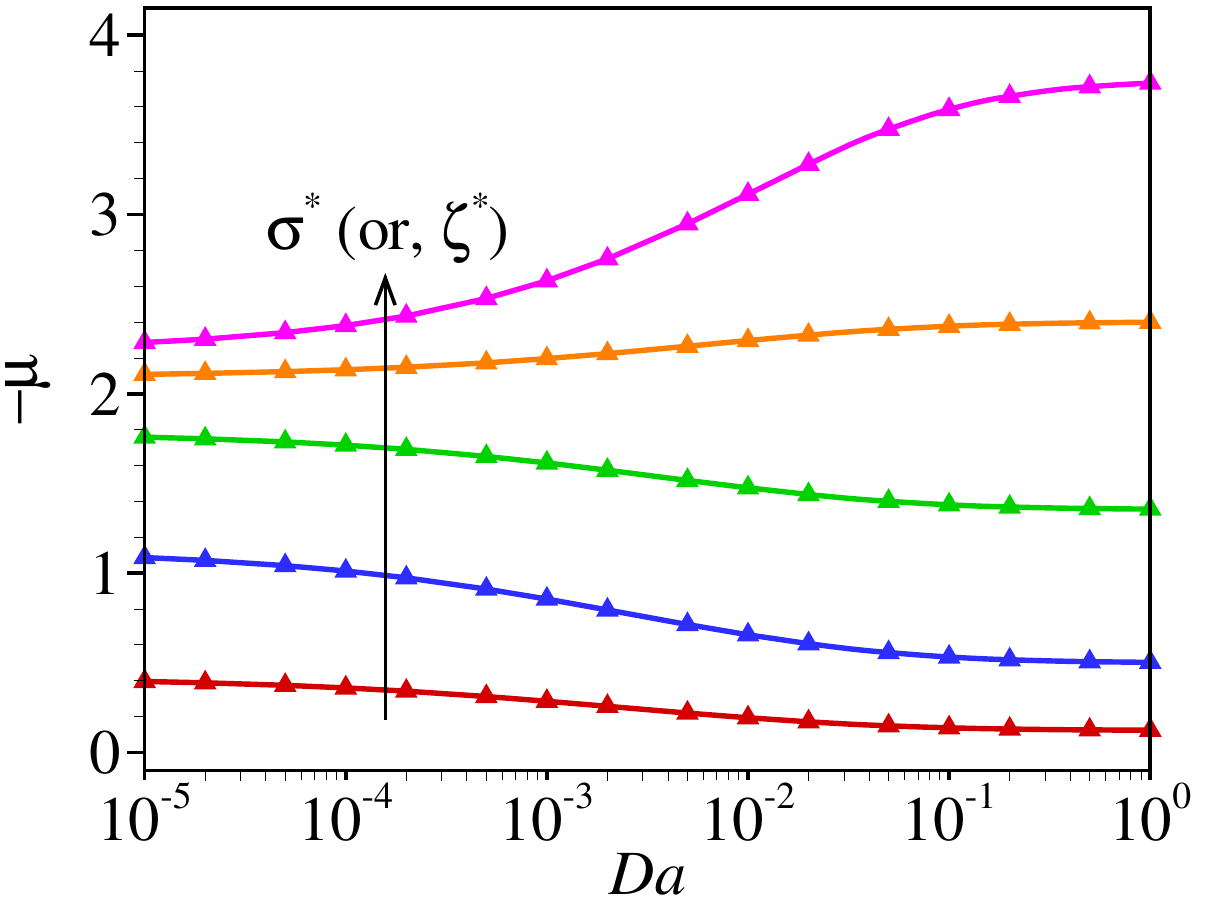}
			\put(2,80){(a)}
		\end{overpic}
	\end{minipage}
	\begin{minipage}[h]{0.3\textwidth}
		\centering
		\begin{overpic}[width=\linewidth]{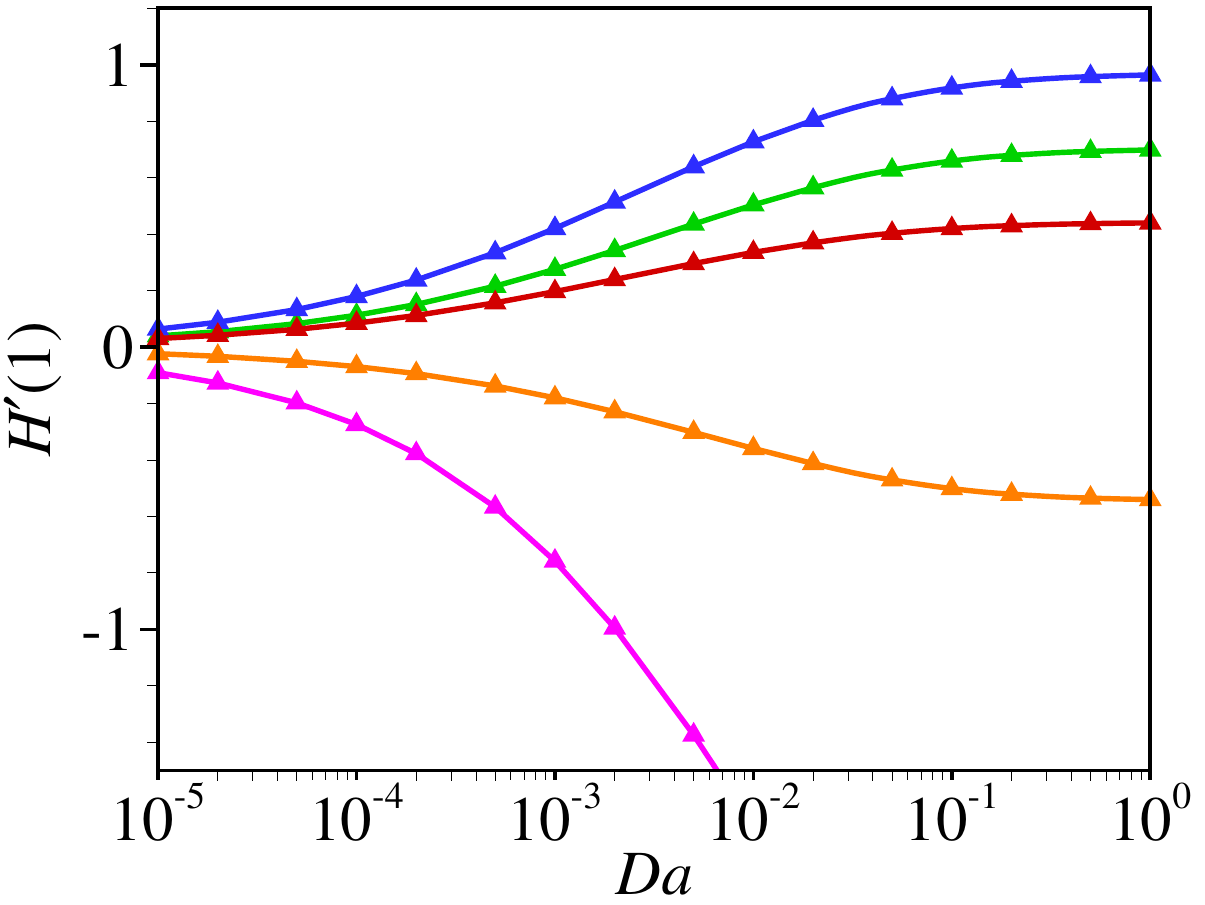}
			\put(2,80){(b)}
		\end{overpic}
	\end{minipage}
	\begin{minipage}[h]{0.38\textwidth}
		\centering
		\begin{overpic}[width=\linewidth]{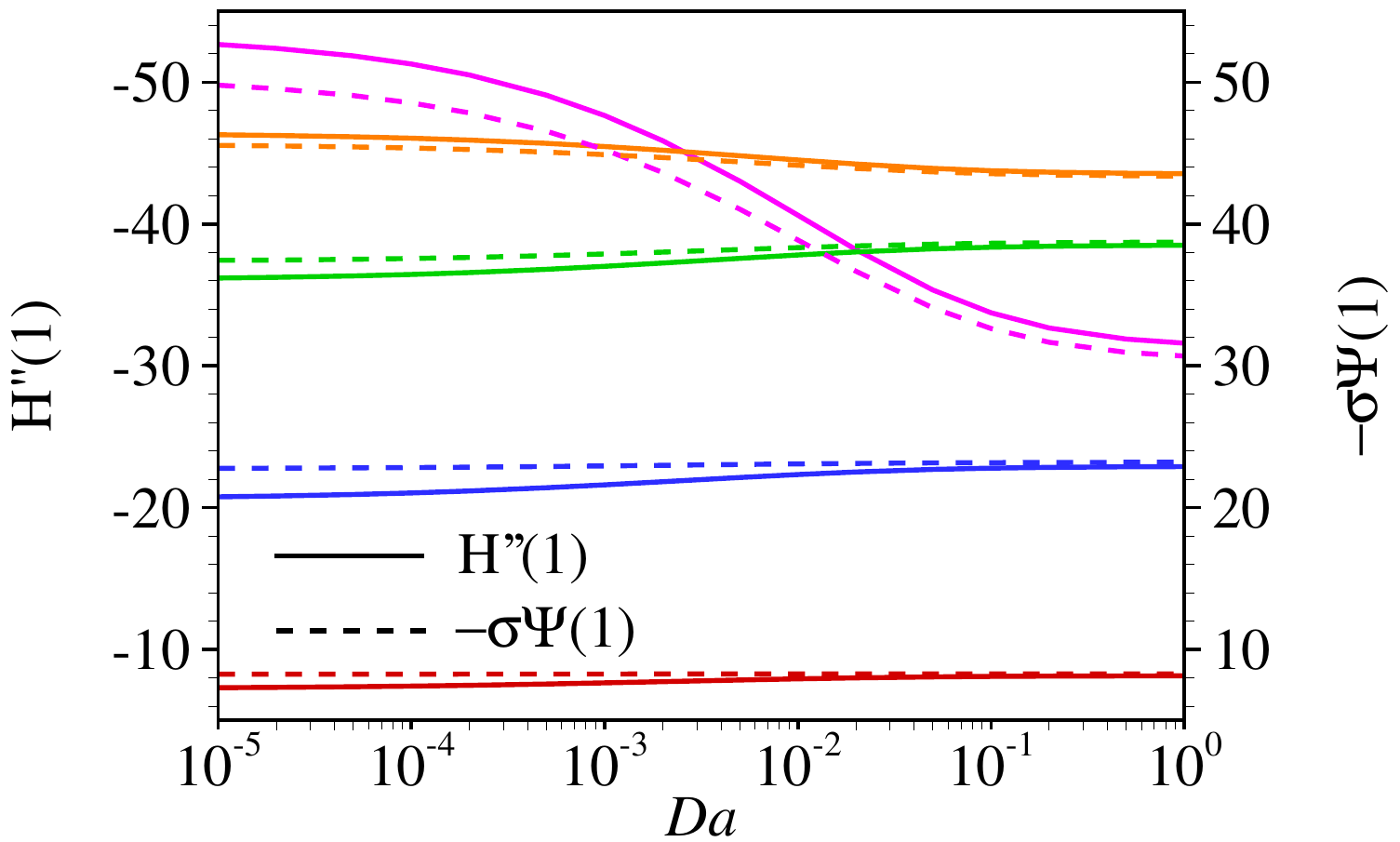}
			\put(2,63){(c)}
		\end{overpic}
	\end{minipage}
	\caption{
		Variation of (a) the electrophoretic mobility $\mu$, (b) the interfacial velocity measure $H'(1)$, and (c) the hydrodynamic and tangential Maxwell stresses as functions of the Darcy number $Da$. The surface charge densities are $\sigma^{*}=-2$ (red lines), $-6$ (blue lines), $-12$ (green lines), $-18$ (yellow lines), and $-24~\mathrm{mC\,m^{-2}}$ (pink lines). The other parameters are fixed at $\kappa a=10$, $\varepsilon_r=0$, and $\eta_r=0.1$ for panels (a)--(c). In panel (c), the left axis represents the tangential hydrodynamic stress, whereas the right axis represents the tangential Maxwell stress.}
	\label{fig:CC}
\end{figure}

We next examine whether the permeability response identified above is controlled only by dielectric polarizability, or whether it can also be altered by changing the surface charge density. To isolate this effect, Fig.~\ref{fig:CC} considers a non-polarizable porous droplet, $\varepsilon_r=0$, at fixed $q=\kappa a=10$ and $\eta_r=0.1$, while varying the imposed surface charge density $\sigma^*$.

Figure~\ref{fig:CC}(a) shows that the effect of permeability depends strongly on the magnitude of the surface charge. At small $|\sigma^*|$, the mobility magnitude, $-\mu$, decreases as $Da$ increases. Thus, making the droplet more permeable weakens the electrophoretic motion relative to the strongly screened case. This behaviour is consistent with the Debye--H\"uckel prediction for a non-polarizable droplet: the internal-circulation correction has the opposite sign to the rigid-particle electrophoretic contribution and therefore reduces the magnitude of the total mobility. As $|\sigma^*|$ is increased, however, this trend weakens and eventually reverses. For sufficiently large surface charge, increasing $Da$ enhances $-\mu$, indicating that the internal circulation now reinforces, rather than opposes, the electrophoretic motion. Hence, for a non-polarizable droplet, a larger surface-charge magnitude can reverse the way in which permeability modifies the mobility.

The origin of this reversal is clarified by Fig.~\ref{fig:CC}(b), where the interfacial-velocity measure $H'(1)$ is plotted against $Da$. For small $|\sigma^*|$, $H'(1)$ is positive. Since the droplet is negatively charged, the rigid-particle contribution to the mobility is negative. Therefore, a positive $H'(1)$ gives a positive correction to the mobility and reduces the magnitude $-\mu$. Physically, this positive interfacial velocity is directed opposite to the electrophoretic motion of the droplet. Increasing $Da$ weakens the Brinkman resistance and amplifies this opposing interfacial motion, which explains the reduction of $-\mu$ at low surface charge. At larger $|\sigma^*|$, however, $H'(1)$ becomes negative. The interfacial motion is then directed along the droplet motion, so the circulation correction has the same sign as the rigid-particle mobility. Increasing permeability now strengthens an assisting circulation and therefore increases the mobility magnitude.

The sign of $H'(1)$ is determined by the signed tangential stress balance at the droplet interface. Figure~\ref{fig:CC}(c) compares the hydrodynamic shear contribution with the tangential Maxwell traction generated by the fixed surface charge. Since the present model assumes uniform surface tension, $\nabla_s\gamma=0$, Marangoni stress is absent; the relevant competition is between hydrodynamic shear and Maxwell traction. With the present sign convention, the tangential Maxwell traction acts in the positive direction, whereas the hydrodynamic shear response associated with the distorted electric double layer acts in the negative direction. When the positive Maxwell traction dominates, the resulting interfacial velocity is positive and opposes the droplet motion. When the negative hydrodynamic shear contribution dominates, the interfacial velocity becomes negative and assists the droplet motion.

This stress-level interpretation explains the surface-charge-induced reversal. At low $|\sigma^*|$, the system remains close to the Debye--H\"uckel regime, and the positive Maxwell traction is sufficiently strong to determine a positive interfacial velocity. The internal circulation then opposes the electrophoretic motion, so increasing permeability reduces $-\mu$. At higher $|\sigma^*|$, nonlinear electric-double-layer polarization becomes important. The redistributed space charge modifies the electrical body force in the diffuse layer and strengthens the negative hydrodynamic shear response. Once this negative hydrodynamic contribution overtakes the direct Maxwell traction, the interfacial velocity reverses sign. The porous droplet then develops an internal circulation that assists electrophoretic migration, and increasing $Da$ enhances the mobility magnitude.

\begin{figure}[H]
	\centering
	\begin{minipage}[h]{0.32\textwidth}
		\centering
		\begin{overpic}[width=\linewidth]{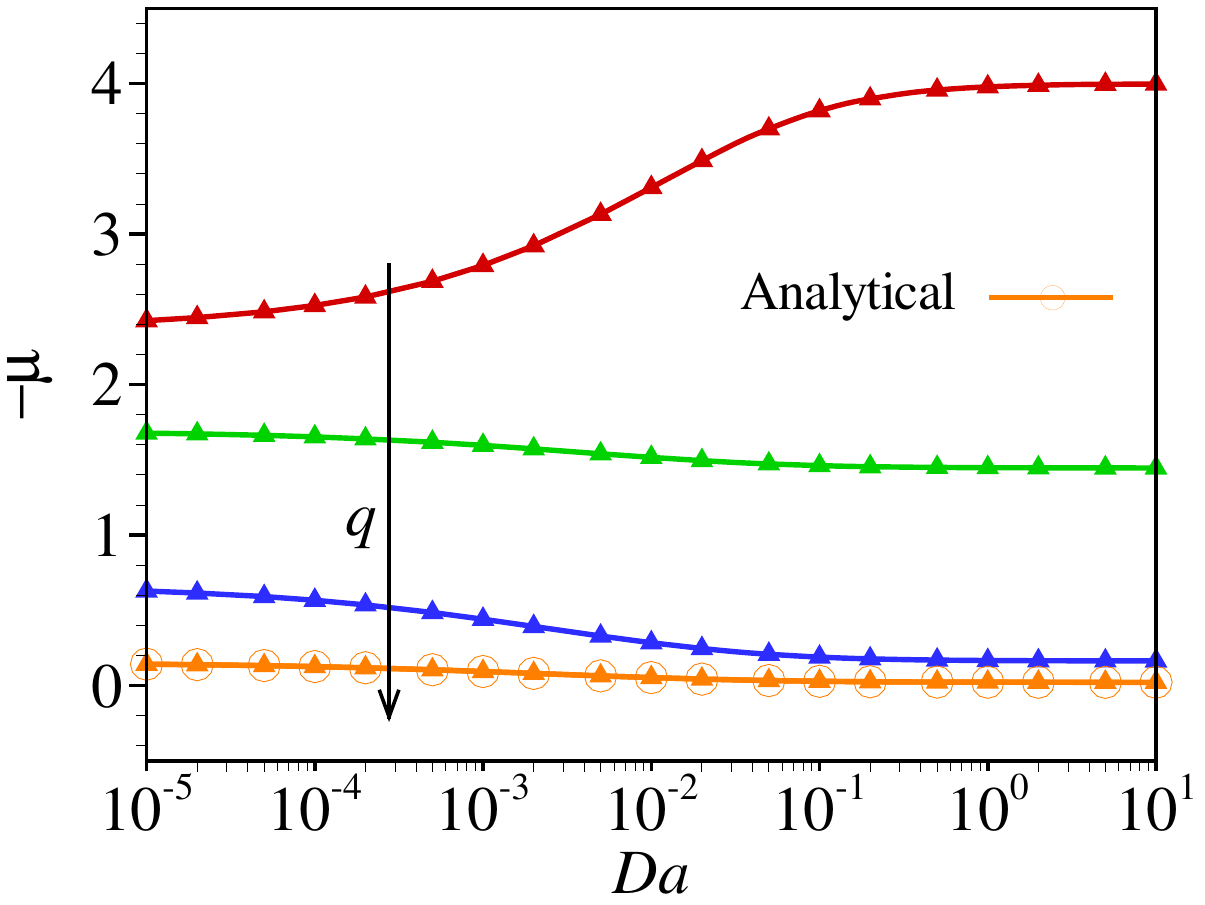}
			\put(2,80){(a)}
		\end{overpic}
	\end{minipage}
	\begin{minipage}[h]{0.32\textwidth}
		\centering
		\begin{overpic}[width=\linewidth]{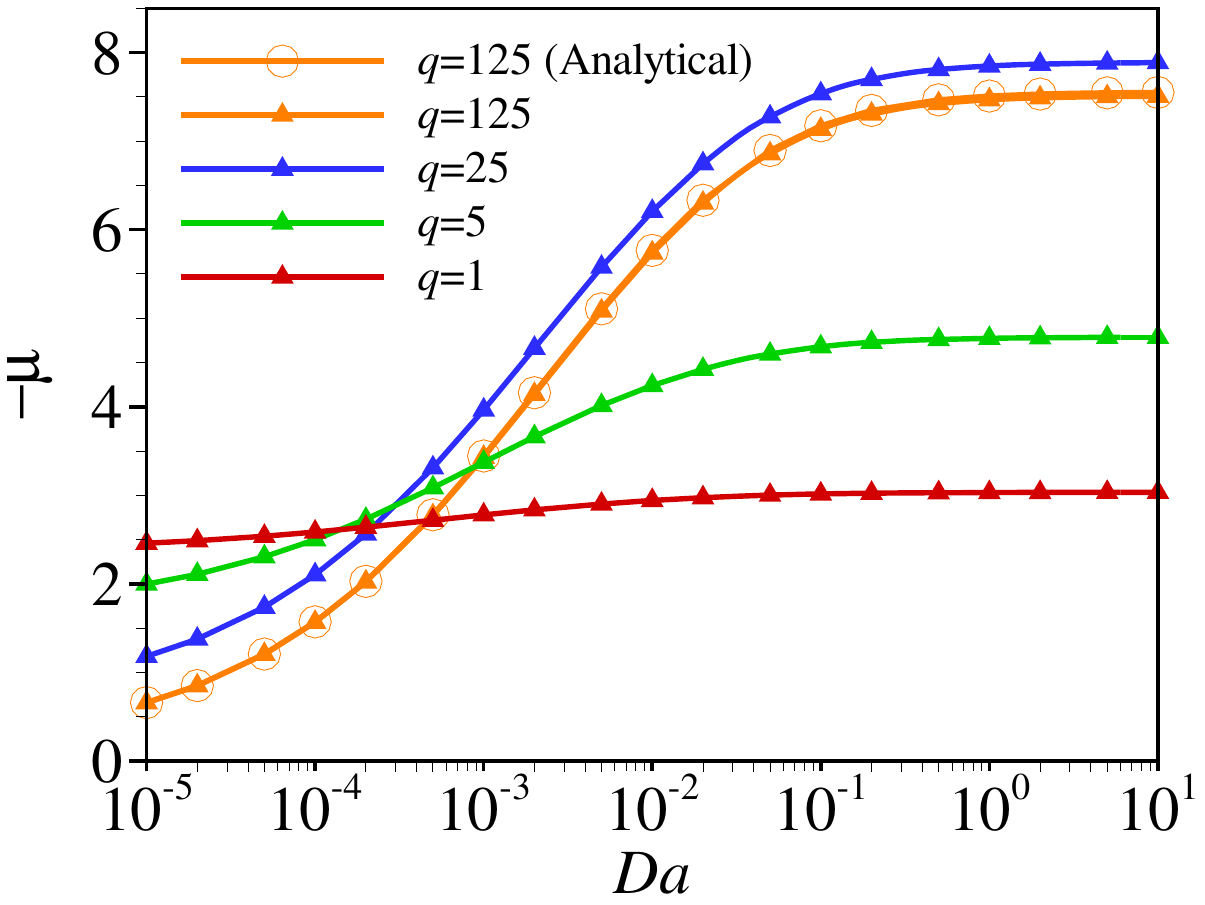}
			\put(2,80){(b)}
		\end{overpic}
	\end{minipage}
	\begin{minipage}[h]{0.32\textwidth}
		\centering
		\begin{overpic}[width=\linewidth]{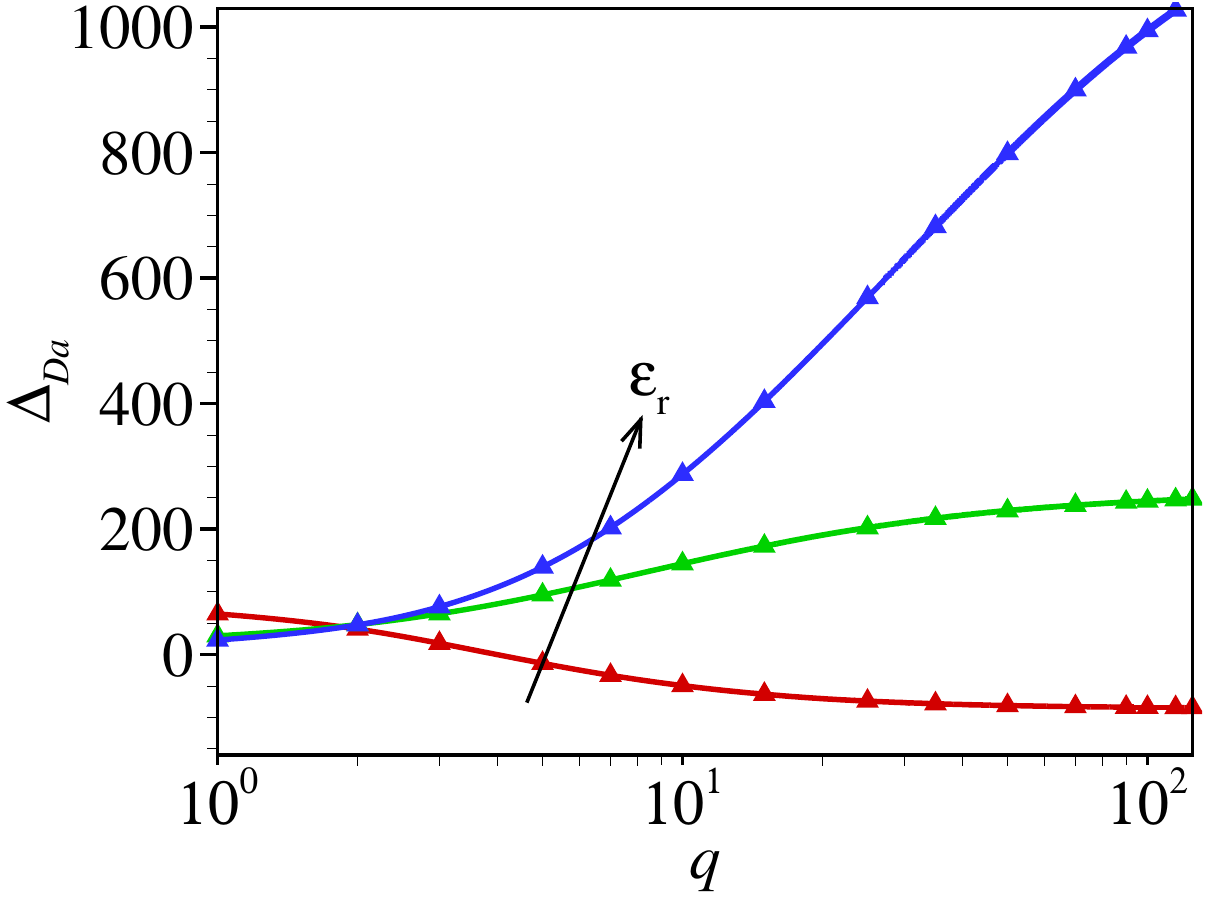}
			\put(2,80){(c)}
		\end{overpic}
	\end{minipage}
	\caption{Variation of the magnitude of the electrophoretic mobility, $-\mu$, as a function of the Darcy number $Da$ for different values of the Debye-layer parameter $q~(=\kappa a=1,5,25,125)$ at (a) $\varepsilon_r=0$ and (b) $\varepsilon_r=10^3$. (c) Variation of the relative Darcy response, $\Delta_{Da}$, with $q$ for different permittivity ratios $\varepsilon_r=0,10,10^3$. The remaining parameters are $\eta_r=0.1$ and $\sigma^{*}=-7~\mathrm{mC\,m^{-2}}$. The orange lines with hollow circles are the analytical expression (\ref{eq:mobility_Smol_general}).}
	\label{fig:DD}
\end{figure}

We now examine how the effect of Darcy number changes with the Debye-layer parameter $q=\kappa a$. Figure~\ref{fig:DD}(a) shows the mobility magnitude $-\mu$ as a function of $Da$ for a non-polarizable droplet, $\varepsilon_r=0$. Two features are evident. First, at a fixed value of $Da$, the mobility generally decreases as $q$ increases. This occurs because the surface charge density $\sigma^*$ is kept fixed. For fixed $\sigma^*$, increasing $q$ decreases the equilibrium surface potential, and therefore weakens the electrokinetic driving responsible for electrophoresis. Secondly, the dependence on $Da$ changes with $q$. For small $q$, increasing $Da$ enhances the mobility magnitude. For larger $q$, however, the trend is reversed and increasing $Da$ reduces $-\mu$. This transition is quantified in Fig.~\ref{fig:DD}(c).

This sign change follows from the interfacial-stress mechanism discussed above. At small $q$, the double layer is relatively thick and, for the present surface charge, nonlinear charge-cloud distortion produces a hydrodynamic shear response that dominates the direct Maxwell traction. The selected interfacial velocity is then directed along the droplet motion, so increasing $Da$ amplifies an assisting circulation and increases $-\mu$. As $q$ increases, the equilibrium surface potential associated with a fixed surface charge decreases and the response approaches the low-potential non-polarizable behaviour. In that regime, the dominant positive tangential Maxwell traction generates an interfacial motion opposite to the droplet motion. Increasing $Da$ then amplifies this opposing circulation, and the mobility magnitude decreases. Thus, the sign change of $\Delta_{Da}$ for $\varepsilon_r=0$ reflects a change in the sign of the interfacial-circulation correction.

Figure~\ref{fig:DD}(b) shows a qualitatively different response for a highly polarizable droplet, $\varepsilon_r=10^3$. In this case, increasing $Da$ enhances $-\mu$ for all values of $q$. The percentage change is also much larger than in the non-polarizable case. As shown in Fig.~\ref{fig:DD}(c), for $\varepsilon_r=10^3$, $\Delta_{Da}$ increases from about $23\%$ at $q=1$ to $76\%$ at $q=3$, $287\%$ at $q=10$, nearly $800\%$ at $q=50$, and more than $1000\%$ for $q\gtrsim100$. The numerical curve for $q=125$ also agrees with the analytical thin-double-layer expression, confirming the strong Darcy-number dependence. This large positive response means that, in the highly polarizable case, permeability has a dramatic influence on the mobility, especially when the double layer is thin.

The reason is that increasing $\varepsilon_r$ weakens the tangential electric field at the droplet surface and therefore weakens the direct tangential Maxwell traction. In the limiting equipotential case, the tangential Maxwell traction vanishes. The interfacial motion is then governed primarily by the hydrodynamic shear response generated by the distorted double layer. For the negatively charged droplet considered here, this response selects a negative interfacial velocity, which assists the electrophoretic motion. Increasing $Da$ reduces Brinkman screening and strengthens this assisting internal circulation. Consequently, the mobility magnitude increases strongly with permeability.

The intermediate case $\varepsilon_r=10$ bridges these two limits. Unlike the non-polarizable case, $\Delta_{Da}$ remains positive over the full range of $q$. It increases from approximately $30\%$ at $q=1$ to $95\%$ at $q=5$, $145\%$ at $q=10$, and about $248\%$ at $q=125$. Thus, even moderate dielectric polarizability is sufficient to shift the system toward permeability-enhanced electrophoresis. However, the amplification is much weaker than for $\varepsilon_r=10^3$, showing that the strength of the Darcy-number response is controlled by how strongly dielectric polarization suppresses the direct Maxwell traction.

\subsection{Mobility transition between clean-droplet and rigid-particle limits}
\begin{figure}[H]
	\centering
	\begin{minipage}[H]{0.42\textwidth}
		\centering
		\begin{overpic}[width=\linewidth]{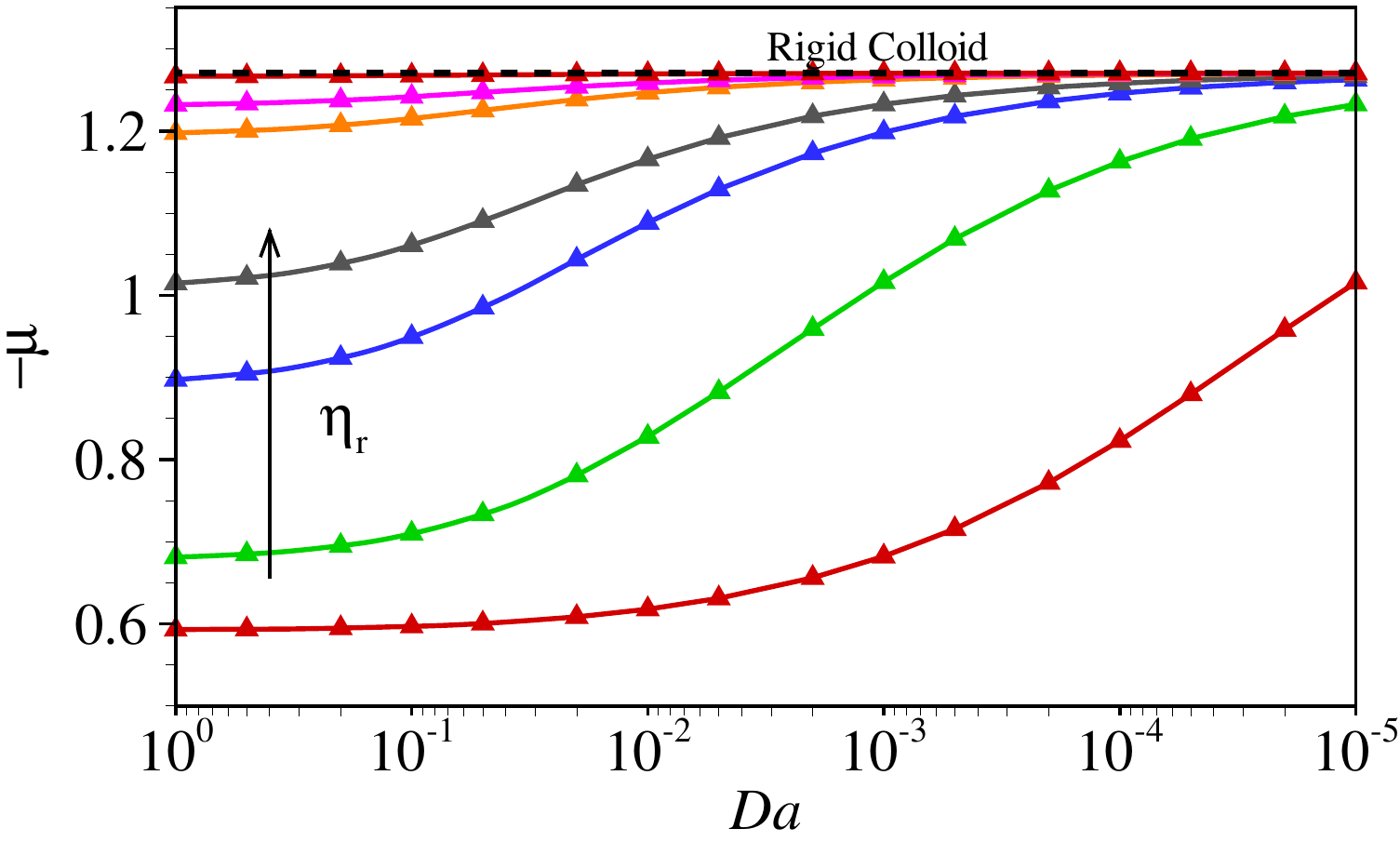}
			\put(2,61){(a)}
		\end{overpic}
	\end{minipage}
	\begin{minipage}[H]{0.42\textwidth}
		\centering
		\begin{overpic}[width=\linewidth]{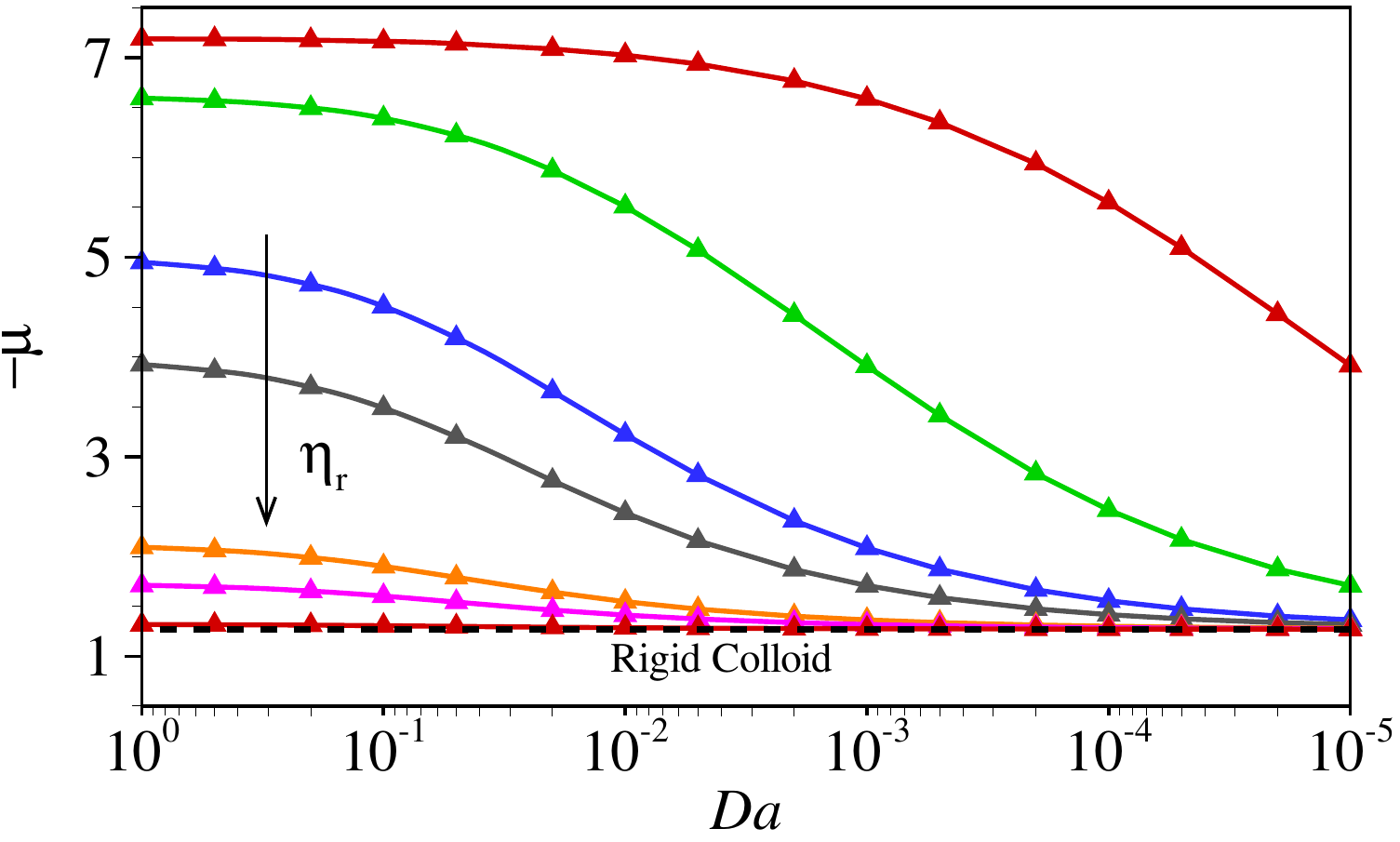}
			\put(2,61){(b)}
		\end{overpic}
	\end{minipage}
	\begin{minipage}[H]{0.42\textwidth}
		\centering
		\begin{overpic}[width=\linewidth]{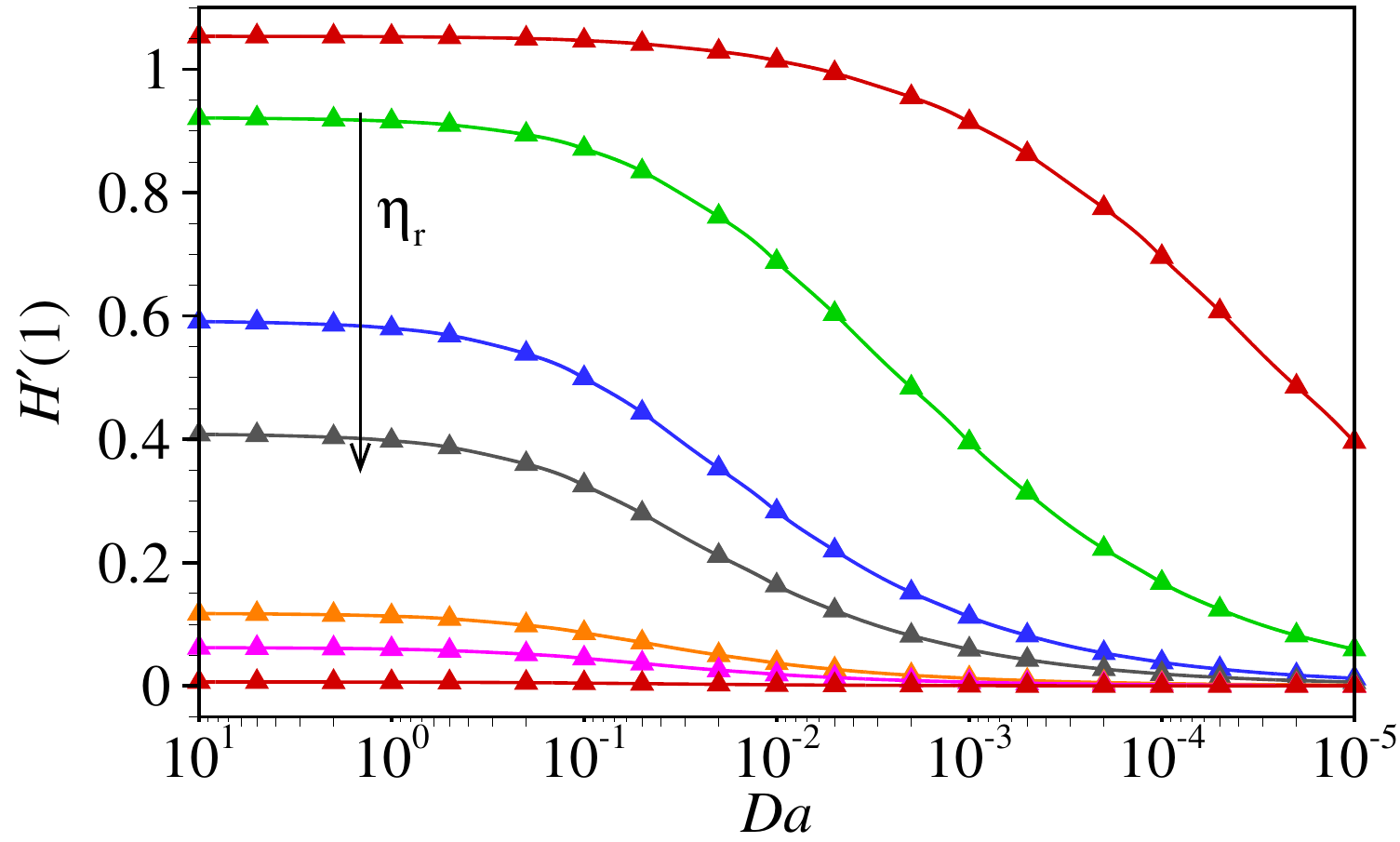}
			\put(2,61){(c)}
		\end{overpic}
	\end{minipage}
	\begin{minipage}[H]{0.42\textwidth}
		\centering
		\begin{overpic}[width=\linewidth]{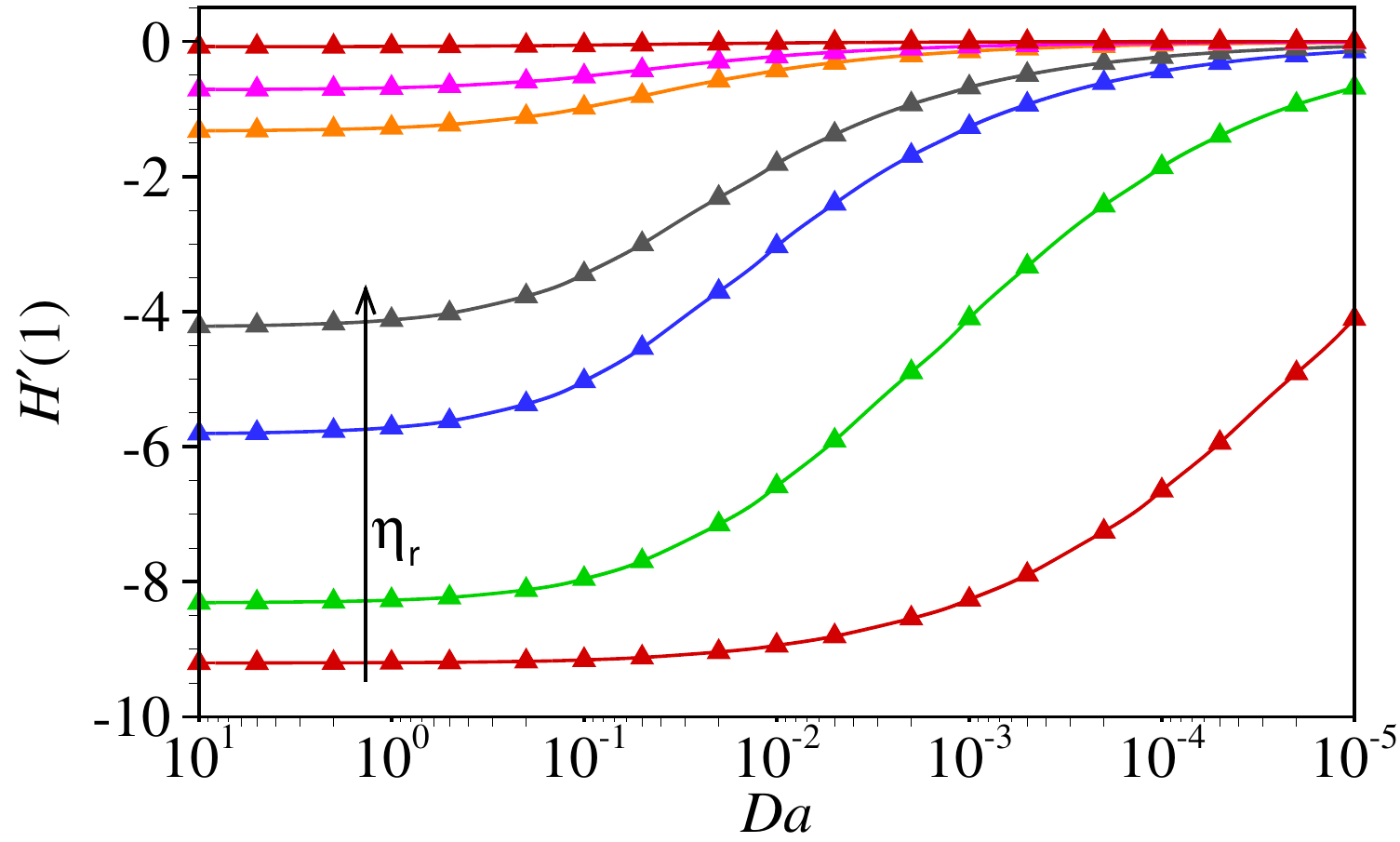}
			\put(2,61){(d)}
		\end{overpic}
	\end{minipage}
	\caption{Variation of the magnitude of the electrophoretic mobility, $-\mu$, and the interfacial velocity measure $H'(1)$ as functions of the Darcy number $Da$ for different viscosity ratios $\eta_r~(=0.01,0.1,0.5,1,5,10,100)$. Panels (a,b) show $-\mu$ for (a) a non-polarizable droplet, $\varepsilon_r=0$, and (b) a highly polarizable droplet, $\varepsilon_r=10^3$, while panels (c,d) show the corresponding variation of $H'(1)$ for (c) $\varepsilon_r=0$ and (d) $\varepsilon_r=10^3$. The remaining parameters are fixed at $\kappa a=10$ and $\sigma^{*}=-7~\mathrm{mC\,m^{-2}}$.}
	\label{fig:EE}
\end{figure}
We now discuss how the Darcy number and viscosity ratio together control the strength of the interfacial motion and the electrophoretic mobility. Figure~\ref{fig:EE}(a) shows the mobility magnitude for a non-polarizable droplet, $\varepsilon_r=0$, as a function of $Da$ for different viscosity ratios. For all $\eta_r$, the curves approach the rigid-colloid mobility as $Da$ decreases. This is expected because the Brinkman resistance becomes large at small $Da$, suppressing the interfacial motion and preventing the exterior electrokinetic forcing from generating internal circulation. As $Da$ increases, the droplet becomes more permeable and the mobility departs from the rigid-particle value. For the non-polarizable case, this departure is downward in $-\mu$, i.e. the mobility magnitude becomes smaller than the rigid-colloid value. The effect is strongest for low-viscosity droplets and becomes progressively weaker as $\eta_r$ increases.

The corresponding interfacial velocity in Fig.~\ref{fig:EE}(c) explains this trend. For $\varepsilon_r=0$, $H'(1)$ is positive and increases with $Da$, especially when $\eta_r$ is small. For the negatively charged droplet considered here, a positive $H'(1)$ gives a mobility correction opposite to the rigid-particle contribution. Thus, increasing $Da$ amplifies an interfacial circulation that opposes the electrophoretic motion, and the magnitude $-\mu$ decreases. Increasing $\eta_r$ has the opposite effect: it increases the internal viscous resistance, reduces $H'(1)$, and therefore brings the mobility back towards the rigid-particle value. Hence, in the non-polarizable case, both small $Da$ and large $\eta_r$ act as hydrodynamic rigidifying mechanisms.

Figures~\ref{fig:EE}(b,d) show the corresponding behaviour for a highly polarizable droplet, $\varepsilon_r=10^3$. The rigid-particle limit is again recovered as $Da$ decreases, but the direction of departure at large $Da$ is now reversed. In Fig.~\ref{fig:EE}(b), increasing $Da$ increases the mobility magnitude above the rigid-colloid value, and this enhancement is most pronounced for low-viscosity droplets. The reason is evident from Fig.~\ref{fig:EE}(d): the interfacial velocity is now negative. Therefore, the circulation correction has the same sign as the rigid-particle electrophoretic contribution and assists the droplet motion. Increasing permeability strengthens this assisting internal circulation, whereas increasing $\eta_r$ suppresses it.

The comparison between panels (a,c) and (b,d) highlights the distinct roles of electrostatics and hydrodynamics. The permittivity ratio determines the sign of the interfacial-velocity mode through the electrostatic and stress-balance problem. Once this sign is selected, $Da$ and $\eta_r$ determine only the magnitude of the hydrodynamic response. A low-viscosity, highly permeable droplet exhibits the strongest liquid-droplet correction because internal circulation is least resisted. In contrast, a highly viscous or poorly permeable droplet has very weak interfacial motion and therefore behaves electrophoretically like a rigid particle, irrespective of whether the selected circulation mode would otherwise assist or oppose the migration.

This behaviour is consistent with the analytical structure of the Debye--H\"uckel solution, where the porous interior enters through an effective internal resistance. Decreasing $Da$ increases the Brinkman screening, while increasing $\eta_r$ increases the viscous resistance of the liquid saturating the porous matrix. Both mechanisms reduce the interfacial velocity and internal circulation. Therefore, the rigid-particle limit is not reached by changing the electric double layer, but by eliminating the hydrodynamic mobility of the interface. Thus, along with the analytical solutions,  the mobility and interfacial-velocity trends in Figure~\ref{fig:EE} demonstrate that a porous liquid droplet forms a continuous hydrodynamic bridge between a clean liquid droplet and a rigid colloid.
\section{Conclusions and outlook}\label{Conclusion}
We have investigated the electrophoresis of a uniformly charged spherical porous liquid droplet whose internal flow is governed by the Brinkman--Debye--Bueche equation. The porous skeleton is taken to be uncharged and impermeable to ions, so that the electric double layer remains confined to the surrounding electrolyte. A regular perturbation expansion in the imposed electric field reduces the full electrokinetic problem to a coupled set of radial ordinary differential equations. The resulting perturbed equations are then solved numerically for arbitrary values of the $\zeta$-potential and Debye layer thickness. 

In the Debye--H\"uckel regime, we derived a closed-form mobility valid for arbitrary double-layer thickness. The analytical result shows that the effect of the porous interior enters through a single Brinkman-screened hydrodynamic resistance. In the highly permeable limit, this resistance reduces to the internal Stokes resistance of a clean liquid droplet, whereas in the low-permeability limit it becomes large enough to suppress interfacial motion. The mobility then approaches the rigid-particle value. Thus, the theory establishes a continuous hydrodynamic transition between clean-droplet and rigid-particle electrophoresis. The analytical solution further identifies the first interfacial-velocity mode as the quantity that controls the departure from rigid-particle electrophoresis. When the interfacial motion is finite, the porous droplet retains a liquid-droplet correction to its mobility. When the interfacial motion is suppressed by Brinkman resistance, the internal circulation disappears and the droplet behaves electrophoretically as a rigid particle.

Numerical solutions beyond the Debye--H\"uckel regime confirm this mechanism and reveal a non-universal role of permeability. Increasing the Darcy number can either reduce or enhance the mobility, depending on the sign of the interfacial-velocity mode. For weakly polarizable droplets, increasing the Darcy number can reduce the mobility magnitude because the induced internal circulation opposes the rigid-particle electrophoretic contribution. For highly polarizable droplets, dielectric polarization weakens the tangential Maxwell traction and can reverse the interfacial-velocity mode; the resulting circulation then reinforces the electrophoretic motion, so that increasing permeability enhances the mobility. A similar reversal can occur at larger surface charge, where nonlinear electric-double-layer polarization modifies the balance between hydrodynamic shear and tangential Maxwell traction.

The dependence on Debye layer thickness shows that this Darcy-number effect persists across a broad electrokinetic range. For non-polarizable droplets, the sign of the permeability response may change with the Debye-layer parameter, whereas for highly polarizable droplets the mobility enhancement can become very large in the thin-double-layer regime. Increasing the viscosity ratio weakens all permeability-induced changes by increasing the effective internal hydrodynamic resistance. In all cases, the rigid-particle limit is recovered when internal circulation is suppressed, either by decreasing the Darcy number or by increasing the droplet viscosity.

Overall, the present study shows that a porous liquid droplet is neither a classical clean droplet nor a rigid colloid. It is a permeability-controlled electrohydrodynamic object whose mobility is governed by the coupling between electric-double-layer forcing, interfacial hydrodynamic shear, electric stresses and Brinkman-screened internal resistance. The framework provides a theoretical basis for tuning the electrokinetic transport of soft porous droplets. 

Natural extensions of the present framework include porosity-dependent effective viscosity, anisotropic permeability, charged or ion-permeable hydrogel interiors, and interfacial transport mechanisms such as mobile surface charge, adsorption--desorption kinetics and Marangoni stresses. For highly charged droplets in concentrated or multivalent electrolytes, finite-ion-size steric effects and ion--ion correlations could also be incorporated, as these effects may alter the electric-double-layer structure and, consequently, the interfacial electrohydrodynamic forcing and electrophoretic mobility. The formulation may also be adapted to related phoretic transport problems, including diffusiophoresis and thermophoretic migration of porous droplets. Since the present theory is based on a weak imposed electric field, it provides a linear-response benchmark for future studies of porous-droplet electrophoresis under strong electric fields, where advective ion transport and nonlinear double-layer polarization become prominent and interfacial deformation may become non-negligible.
\bibliography{reference}
\end{document}